\documentclass[12pt]{amsart}
\usepackage{geometry}                % See geometry.pdf to learn the layout options. There are lots.
\usepackage{graphicx}
\usepackage{amssymb}
\usepackage{epstopdf}
\usepackage{mathtools}
\usepackage{cite}
\usepackage{enumitem}
\usepackage{microtype}

\newtheorem{theorem}{Theorem}
\newtheorem{proposition}{Proposition}

\newtheorem{definition}{Definition}
\newtheorem{lemma}{Lemma}

\newtheorem{remark}{Remark}
\newtheorem{example}{Example}

\newtheorem*{conjecture}{Conjecture}
\newtheorem*{remarknn}{Remark}

\newcommand{\C}{\mathbb{C}}
\newcommand{\Z}{\mathbb{Z}}
\newcommand{\N}{\mathbb{N}}

\newcommand{\ord}[1]{\operatorname{ord}(#1)}
\newcommand{\mc}[1]{\begin{pmatrix*}[c]#1\end{pmatrix*}}

\newcommand{\uu}{\mathfrak u}
\newcommand{\vv}{\mathfrak v}
\newcommand{\ds}{\displaystyle}

\newcommand{\p}{\partial}
\newcommand{\be}{\begin{equation}}
\newcommand{\ee}{\end{equation}}

\allowdisplaybreaks

\title{The symmetry approach to integrability: recent advances}
\author{Rafael Hern\'andez Heredero}
\address{Depto.~de Matem\'atica Aplicada a las TIC,
 Universidad Polit\'ecnica de Madrid,
C.~Nikola Tesla s/n. 28031 Madrid. Spain}
\email{rafahh@etsist.upm.es}
\author{Vladimir Sokolov}
\address{Landau Institute for Theoretical Physics, 
142432 Chernogolovka (Moscow region), Russia}
\curraddr{Universidade Federal do ABC, 09210-580 Sao Paulo, Brazil}
\email{vsokolov@landau.ac.ru}
\begin{document}
\begin{abstract}
We provide a concise introduction to the symmetry approach to integrability. Some results on integrable evolution and systems of evolution equations are reviewed. Quasi-local recursion and Hamiltonian operators are discussed. We further describe non-abelian integrable equations, especially matrix (ODE and PDE) systems. Some non-evolutionary integrable equations are studied using a formulation of formal recursion operators that allows to study non-diagonalisable systems of evolution equations.
\end{abstract}

\maketitle

\section{Introduction}

The symmetry approach to the classification of integrable PDEs  is being developed since 1979 by: A.~Shabat, A.~Zhiber,  N.~Ibragimov,  A.~Fokas,
V.~Sokolov, S.~Svinolupov, A.~Mikhailov,  R.~Yamilov, V.~Adler,
P.~Olver, J.~Sanders, J.P.~Wang, V.~Novikov, A.~Meshkov, D.~Demskoy, H.~Chen, Y.~Lee, C.~Liu, I.~Khabibullin,   B.~Magadeev, R.~H.~Heredero,   V.~Marikhin,  M.~Foursov, S.~Startcev, M.~Balakhnev, and others. It is very efficient for PDEs with two independent variables and, under additional assumptions, it can be applied for ODEs.
 
The basic definition of the symmetry approach is the following. 
 
\begin{definition} A differential equation is integrable if it possesses infinitely many higher infinitesimal symmetries.
\end{definition}

This definition is a priori reasonable: linear equations have infinitely many higher symmetries and all known integrable equations are related to linear equations by some transformations. These transformations produce symmetries on the integrable nonlinear equations coming from symmetries of the corresponding linear equations.

Requiring the existence of higher symmetries is a powerful method to find all integrable equations  from a prescribed class of equations. The first classification result in the frame of the symmetry approach was the following:
\begin{theorem}[\hspace{1sp}\cite{zibshab1}]\label{ZS} 
A nonlinear hyperbolic equation
of the form
\[
u_{xy}=F(u)
\]
possesses higher symmetries iff $($up to scalings and shifts$)$
\[
F(u)=e^{u}, \quad F(u)=e^u+e^{-u}, \quad \mbox{or} \quad F(u)=e^u+e^{-2u}.
\]
\end{theorem}

There are several reviews~\cite{sokshab,MikShaYam87,int2,MikShaSok91,ASY,AMS} devoted to the symmetry approach (see also~\cite{ibrag,Fokas,KLV,Olv93}). In this report we mostly concentrate on results not covered in those papers and books. Due to lack of space, we prefer to illustrate some significant ideas by examples and to give informal but constructive definitions, and refer to the cited reviews (e.g.~\cite{Olv93,MikShaSok91,int2}) for more detailed and rigorous developments. With respect to citations of original sources, we usually cite general reviews where the references can be found. 

\subsection{Infinitesimal symmetries}\label{sec1.1}

Consider a dynamical system of ODEs
\begin{equation}
\frac{dy^{i}}{dt} = F_{i}(y^1,\ldots ,y^n),\qquad i=1, \ldots ,n
\, .  \label{dynsys}
\end{equation}
 
\begin{definition} The dynamical system 
\begin{equation}
\frac{dy^{i}}{d\tau} = G_{i}(y^1,\ldots ,y^n),\ \ \ \ i=1, \ldots
,n  \,   \label{dynsym}
\end{equation}
is called an infinitesimal symmetry of~\eqref{dynsys} iff~\eqref{dynsys} and~\eqref{dynsym}
are compatible.
\end{definition}
Compatibility means that $X Y-Y X=0$, where 
\begin{equation}\label{ODEfield} X=\sum F_i \frac{\partial}{\partial y^i},  \quad {\rm and} \quad Y=\sum G_i \frac{\partial}{\partial y^i}.
\end{equation}  

Consider now an evolution equation
\begin{equation}\label{eveq}
u_t=F(u, u_x,  u_{xx}, \dots , u_n), \qquad u_i=\frac{\partial^i
u}{\partial x^i}.
\end{equation}
A higher (or generalized) infinitesimal symmetry of~\eqref{eveq} is an evolution equation
\begin{equation}\label{evsym}
u_{\tau}=G(u, u_x,  u_{xx}, \dots , u_m), \qquad m > 1
\end{equation}
that is compatible with~\eqref{eveq}. 
\begin{remarknn} Infinitesimal symmetries~\eqref{evsym} with $m \le 1$ correspond to one-parameter groups of point or contact transformations~\cite{Olv93}, the so called {\it classical symmetries}.
\end{remarknn}

Compatibility of~\eqref{eveq} and~\eqref{evsym} means that 
\[
\frac{\partial}{\partial t}\frac{\partial u}{\partial \tau}=\frac{\partial}{\partial \tau}\frac{\partial u}{\partial t},
\]
where the partial derivatives are calculated in virtue of~\eqref{eveq} and~\eqref{evsym}. The compatibility condition can be rewritten as~$G_*(F)=F_*(G)$ or 
\begin{equation}\label{DFsym}
D_t(G)-F_*(G)=0.
\end{equation}
Here and below for any function $ a(u, u_1, \dots )$ we denote  
\[
a_* \stackrel{\rm def} {=} \sum_k \frac{\partial a}{\partial u_k} D^k \,,
\]
where 
\[
D= \sum_{i=0}^\infty u_{i+1} \frac{\partial}{\partial u_i}
\]
is the total $x$-derivative.

The concept of infinitesimal symmetry is connected with the  procedure of linearization of a differential equation. For an arbitrary partial differential equation
\[
Q (u, u_x, u_t, \dots) = 0
\]
consider a linear equation
\[
0 = \left.\frac {\p} {\p \varepsilon} \Big (Q (u + \varepsilon \varphi, u_x + \varepsilon \varphi_x, u_t + \varepsilon \varphi_t, \dots \Big) \right| _ {\varepsilon = 0} \stackrel {\rm def} {=} \mathcal{L}_Q (\varphi),
\]
where the linear differential operator
\[
\mathcal{L}_Q = \frac {\p Q} {\p u} + \frac {\p Q} {\p u_ {x}} \, D + \frac {\p Q} {\p u_ {t }} \, D_t + \dots
\]
is called the \emph{linearization operator} for the equation $ Q = 0. $  Here $D$ and $D_t$ are the total derivatives~\cite{Olv93} with respect to $x$ and $t$, respectively.

In the case of an evolution equation~\eqref {eveq}, we have~$Q=u_t-F$ and
\[
\mathcal{L}_Q = D_t - F_ *.
\]
Formula~\eqref{DFsym} means that the symmetry generator, $G$,  is an element of the kernel of the linearization operator. This is a definition of infinitesimal symmetry which is directly applicable to the case of non-evolution PDEs.

For a more rigorous definition of symmetries of evolution equations in terms of evolution vector fields, see~\cite{Olv93,int2}.

\subsection{Examples}

In this section we present several examples of polynomial evolution equations and their symmetries, and also how the existence of a symmetry allows to find integrable equations. 

\begin{example} Any equation of the form~\eqref {eveq} has the classical symmetry $ u _ {\tau} = u_x, $ which corresponds to the group of displacement parameters $ x \to x + \lambda $.
\end {example}

\begin {example} \label{Example1.4}
For all $ m $ and $ n $, the equation $u_{\tau} = u_m $ is a symmetry of the linear equation $ u_t = u_n $. Symmetries with different 
$ m $ are compatible with each other. Thus, we have an infinite {\it hierarchy} of evolution linear equations such that each of the equations is a symmetry for all the others.
\end {example}

\begin{example} \label{Example1.5}
The Burgers equation
\be \label{burgers}
u_t = u_ {xx} +2 u u_x
\ee
has a third-order symmetry
\[
u_{\tau} = u_{xxx} +3 u u_{xx} + 3 u_x^2 + 3 u^2 u_x.
\]
\end{example}

\begin{example} \label{Example1.6}
The simplest higher symmetry of the Korteweg-de Vries (KdV) equation
\begin{equation} \label{kdv} 
u_t=u_{xxx}+ 6\,uu_x
\end{equation}
is 
\begin{equation}\label{kdvsym}
u_{\tau}=u_{5}+10\, u u_{3}+20\, u_1 u_2+30\, u^2 u_1.
\end{equation}
\end{example}

Consider equations of the form 
\begin{equation} \label{kdv5}
u_t = u_5 + a_1\, u u_3 + a_2\, u_1 u_2 + a_3\, u^2 u_1,
\end {equation}
where $ a_i$ are constants.
Let us find all  equations~\eqref {kdv5} admitting a symmetry  
\[
u _ {\tau} = u_7 + c_1 u u_5 + c_2 u_1 u_4 + c_3 u_2 u_3 + c_4 u^2 u_3 + c_5 u u_1 u_2
+ c_6 u_1^3 + c_7 u^3 u_1.
\]

The left hand side of the  compatibility condition~\eqref {DFsym}  is a  polynomial $P$ in the variables $ u_1, \dots, u_{10} $.  There are no linear terms in $ P $. Equating to zero  the coefficients of quadratic terms, we find,
\[ c_1 = \frac{7}{5} a_1, \qquad c_2 = \frac{7}{5} (a_1 + a_2),
\qquad c_3 = \frac{7}{5} (a_1 + 2 a_2).
\]
The vanishing conditions for cubic part allow us to express
$ c_4, c_5 $ and $ c_6 $ in terms of 
$ a_1, a_2, a_3 $. In addition, it turns out that
\[ a_3 = - \frac {3}{10} a_1^2 + \frac {7}{10} a_1 a_2- \frac {1}{5} a_2^2.
\]
Fourth degree terms lead to a formula expressing $ c_7 $ in terms of
$ a_1, a_2 $ and to the basic algebraic relation
\[ (a_2-a_1) (a_2-2 a_1) (2 a_2-5 a_1) = 0 
\]
between the coefficients $ a_1 $ and $ a_2 $. Solving this equation, we find that up to a scaling 
  $ u \rightarrow
\lambda u $ there are only four integrable cases: the linear equation
$ u_t = u_5, $ equations
\begin{equation} \label{sk}
u_t = u_5 + 5 u u_3 + 5 u_1 u_2 + 5 u^2 u_1,
\end {equation}
\begin{equation} \label{kk}
u_t = u_5 + 10 u u_3 + 25 u_1 u_2 + 20 u^2 u_1,
\end {equation}
and~\eqref {kdvsym}.
In each of these cases, the terms of the fifth and sixth degrees in the defining equation are canceled automatically.
The equations~\eqref{sk} and~\eqref {kk} are well known~\cite{SawKot74, Kau80}.

Subsection~\ref{SectionIntegrabilityConditions} deals with an advanced version of symmetry test, where there is no requirement for the polynomiality of the right-hand side of equation~\eqref{eveq}, as well as fixing the order of symmetry. Only the existence of an infinite hierarchy of symmetries is required. 
A similar approach is being developed for equations with local higher conservation laws.

\subsection{First integrals and local conservation laws}\label{conslaw}

In the ODE case the concept of a first integral (or integral of motion) is one of the basic notions. A function $f(y^1,\dots,y^n)$ is  a {\it first integral} for the system~\eqref{dynsys} if its value does not depend on $t$ for any solution $\{y^1(t),\dots,y^n(t)\}$ of~\eqref{dynsys}.  Since
\[
\frac{d}{dt} \Big( f(y^1(t),\dots,y^n(t))\Big) = X(f),
\]
where the vector field $X$ is defined by~\eqref{ODEfield}, from the algebraic point of view a first integral is a solution of the first order PDE
\[
X \Big(f(y^1,\dots,y^n)\Big)=0.
\]
In the case of evolution equations of the form~\eqref{eveq} the concept of integral of motion is substituted by that of {\it local conservation law}: a pair of functions
$\rho(u,u_x,...)$ and $\sigma(u,u_x,...)$ such that 
\begin{equation}\label{conlaw}
 \frac{\partial}{\partial t}\Big(\rho(u, u_x,\dots, u_p) \Big)= \frac{\partial}{\partial x}\Big(\sigma(u, u_x,\dots, u_q) \Big)
\end{equation}
for any solution $u(x,t)$ of~\eqref{eveq}.
 The functions $ \rho $ and $ \sigma $ are called {\it density} and {\it flow} of the conservation law~\eqref{conlaw}, respectively.  

For soliton-type solutions,  which are decreasing with derivatives as  $ x \to \pm \infty $, we obtain
\[
 \frac{\partial}{\partial t} \int _ {- \infty} ^ {+ \infty} \rho \, dx = 0
\]
for any polynomial density $\rho$ with a constant free term. This justifies the name.
{\it conserved density} for the function $ \rho $. Similarly, if $ u (x, t) $ is periodic in $ x $ with a period$ L $, then the value of the functional $\int_0^L \rho \, dx $ on the solution $u$ does not depend on time, so it is an integral of motion for equation~\eqref{eveq}.

\begin{example} \label{Example1.7}
The functions
\[
\rho_1=u,\qquad \rho_2=u^2,\qquad \rho_3=-u_x^2+2u^3
\]
are conserved densities for the Korteweg-de Vries equation~\eqref{kdv}.
\end{example}

\begin{example} \label{Example1.8} For any $n$ the function $\rho_n = u_n^2$ is a conserved density for the linear equation 
\begin{equation} \label{ln3}
u_t = u_3.
\end {equation}
\end{example}

\section{The symmetry approach to integrability}\label{sec:sai}

The symmetry approach to the classification of integrable PDEs with two independent variables is based on the existence of higher symmetries and/or local conservation laws. 

  In the terminology by F. Calogero, an equation is \emph{$S$-integrable} if it has infinitely many higher symmetries \emph{and} conservation laws.  And it is~\emph{$C$-integrable}  if it has infinitely many  higher symmetries but only a finite number of higher conservation laws.

\begin{proposition}[\hspace{1sp}\cite{AbGal1}, Theorem 29 in~\cite{int2}]\label{nocons1}
Scalar evolution equations~\eqref{eveq} of even order $n =  2 k$ cannot possess infinitely many  higher local conservation laws.
\end{proposition}

Typical examples of \emph{$S$-integrable} and \emph{$C$-integrable} are the KdV-equation~\eqref{kdv} and the Burgers equation~\eqref{burgers}, respectively.

\begin{remarknn} Usually, the inverse scattering method can be applied to $S$-integrable equations  while $C$-integrable equations can be reduced to linear equations by differential substitutions. However, to eliminate obvious exceptions like the linear equation~\eqref{ln3}, we need to refine the definition of $S$-integrable equations (see Definition \ref{clar}). 
\end{remarknn}

There are two types of classifications obtained with the symmetry approach:  a ``weak'' version, with equations admitting conservation laws and symmetries ($S$-integrable equations), and a ``strong'' version related only to symmetries, containing both $S$-integrable and $C$-integrable equations.

\subsection{Description of some classification results}
\label{subs:someclass}

\subsubsection{Hyperbolic equations.}

The first classification result using the symmetry approach  was formulated in Theorem~\eqref{ZS}, and concerned hyperbolic equations.
Nevertheless, to fully classify more general integrable hyperbolic equations remains an open problem. Some partial results were obtained in~\cite{meshsokHyp,zib}. 

\begin{example} The following equation
\[
u_{xy}=S(u) \sqrt{\vphantom{u_y^2}u_x^2+1}\sqrt{u_y^2+1}, \qquad {\rm where} \qquad 
S''-2 S^3+c \, S=0,
\]
is integrable. 
\end{example}
For hyperbolic equations~$u_{xy}=\Psi(u, u_{x}, u_{y})$ the symmetry approach assumes the existence of both $x$-symmetries of the form~$u_{t}=A(u, u_{x}, u_{xx}, \dots, )$
and $y$-symmetries of the form~$
u_{\tau}=B(u, u_{y}, u_{yy}, \dots, )$, as it happens with the famous integrable sin-Gordon equation~$u_{xy}=\sin{u}$.
In~\cite{meshsokHyp} a classification was given assuming that both $x$ and $y$-symmetries are integrable evolution equations of third order. In the survey~\cite{zibsok} the reader can find further results on integrable hyperbolic equations.

\subsubsection{Evolution equations.}

For evolution equations of the form~\eqref{eveq}, some necessary   conditions for the existence of higher symmetries not depending on symmetry order were found in~\cite{ibshab, sokshab} (cf.~Subsection 2.2). It was proved in~\cite{soksvin1} that the same conditions hold if the equation~\eqref{eveq} admits infinitely many
local conservation laws. In fact, the conditions for conservation laws are stronger (see Theorem \ref{contriv}) than the conditions for symmetries.

\paragraph{Second order equations.}
All nonlinear integrable equations of the form
\[
u_t=F(x,\,t,\, u,\,u_1,\,u_2)
\]
were listed in~\cite{svin4} and~\cite{soksvin}.
The answer is:
\begin{align*}u_t&=u_2+2 u u_x+h(x),\\
u_t&=u^2 u_2-\lambda x u_1+\lambda u, \\
u_t&=u^2 u_2+\lambda u^2,\\
u_t&=u^2 u_2-\lambda x^2 u_1+3 \lambda x u.
\end{align*}
This list is complete up to contact transformations of the form
\begin{gather*}
\hat t=\chi(t),  \qquad
\hat x=\varphi(x,u,u_1), \qquad \hat u=\psi(x,u,u_1), \\
\hat u_i=\left(\frac{1}{D(\varphi)} D\right)^i(\psi),\qquad
D(\varphi) \frac{\partial \psi}{\partial u_1}=D(\psi) \frac{\partial
\varphi}{\partial u_1}.
\end{gather*}
The first three equations of the list possess local symmetries and form a list
obtained in~\cite{svin4}. The latter equation has so called weakly non-local symmetries (see~\cite{soksvin}). According to Proposition \ref{nocons1} all these equations are $C$-integrable. 
They are related to the heat equation $v_t=v_{xx}$ 
by differential substitutions of Cole-Hopf type~\cite{SvSok26}.

\paragraph{Third order equations.} 
A first result of the ``weak'' type for equations~\eqref{eveq} is the following:
\begin{theorem}[\hspace{1sp}\cite{soksvin1}] A complete list $($up to ``almost invertible'' transformations {\rm~\cite{SvSok26}}$)$
of equations of the form 
\begin{equation} \label{vvs}
u_{t}=u_{xxx}+f(u, u_x, u_{xx})
\end{equation}
with an infinite sequence of conservation laws can be
written as:
\begin{align}
u_t&=u_{xxx}+6u \,u_{x},\nonumber\\
u_t&=u_{xxx}+u^2u_{x},\nonumber\\
u_t&=u_{xxx}-\frac{1}{2}u_{x}^3+(\alpha e^{2u}+\beta
e^{-2u})u_{x},\nonumber\\
u_{t}&=u_{xxx}-\frac{1}{2} Q'' \,u_x + \frac{3}{8} 
\frac{(Q-u_x^2)_{x}^2}{u_x\,(Q-u_x^2)},\nonumber\\
\label{KN}
u_t&=u_{xxx}-\frac{3}{2}\,\frac{u^{2}_{xx}+Q(u)}{u_{x}},\qquad\text{with~$\,\,Q^{(\rm v)}(u)=0$}.
\end{align}
\end{theorem}
For the ``strong'' version of this Theorem see~\cite{soksvin2,meshsok}. 

More general integrable third order equations of the form~$u_t=F(u,u_x,u_{xx},u_{xxx})$ admit three possible types of $u_{xxx}$-dependence~\cite{MikShaSok91}:
\[\text{1)}\ 
u_t=a \,u_{xxx}+b,\quad\text{2)}\ 
u_t=\frac{a}{(u_{xxx}+b)^2}+c,\quad\text{and 3)}\ 
u_t=\frac{2 a\, u_{xxx}+b}{\sqrt{a \,u_{xxx}^2+b\, u_{xxx}+c}}+d,
\]
where the functions $a,b,c$ and $d$ depend on $u,u_x,u_{xx}$. 
A complete classification of integrable equations of such type is not finished yet~\cite{hss1,her}, but there is the following insight.
\begin{conjecture}
All integrable third order equations are related to the KdV equation or to the Krichever-Novikov equation~\eqref{KN} by differential substitutions  of Cole-Hopf and Miura type~\cite{cvsokyam}.
\end{conjecture}
 
\paragraph{Fifth order equations.}
All equations of the form
\[
u_t=u_5+F(u, u_x, u_2, u_3, u_4),
\]
possessing higher conservation laws were found in~\cite{DSS}. 

The list of integrable cases contains well-known equations and several new equations like
\begin{multline*}
u_t=u_{5}+5 (u_{2}-u_1^2+\lambda_1 e^{2u}-
\lambda_2^2 e^{-4u})\,u_3 -5 u_1 u_{2}^2
\\
{} +15 (\lambda_1 e^{2u}\, u_{3}+ 
 4 \lambda_2^2 e^{-4u})\, u_1
u_{2}+ \,u_1^5 -90 \lambda_2^2 e^{-4u}\, u_1^3+ 5(\lambda_1 e^{2u}-\lambda_2^2
e^{-4u})^2\, u_1.
\end{multline*}
The ``strong'' version of this classification result appears in~\cite{meshsok}.

The problem of classifying integrable equations \begin{equation}
u_t=u_n+F(u, \, u_x, \, u_{xx}, \dots, u_{n-1}), \qquad u_i=\frac{\partial^i u}{\partial x^i} \label{scalar}
\end{equation} 
with arbitrary $n$ seems to be far of being solved. Nevertheless, there are some clues to conclude that the only relevant classifications are those with~$n=2,3,5$. Each integrable equation together with all its symmetries form a hierarchy of integrable equations. As a rule, the members of a hierarchy also commute between themselves, i.e.~each equation of the hierarchy is a higher symmetry for all others~(cf.~\cite{sok1} for more details).
In the case of equations~\eqref{scalar} polynomial and homogeneous, it was proved in~\cite{sw,ow} that the corresponding hierarchy contains an equation of second, third, or fifth order. It seems quite plausible that this fact could be extended to the general, non-polynomial case~\eqref{scalar}.  

Further references on the classification of scalar evolution equations are the reviews~\cite{sokshab, MikShaYam87, MikShaSok91, int2, meshsok} and papers~\cite{CheLeeLiu79,fokas,Kap82,AbeGal83}.

\subsubsection{Systems of two equations.}

In~\cite{MikSha85,MikSha86} necessary  conditions of integrability were generalized to the case of systems of evolution equations. Computations become involved and the most general classification problem solved~\cite{MikSha85,MikSha86,MikShaYam87} is that of all $S$-integrable systems  systems of the form
\begin{equation}\label{sys2}
u_t=u_{2}+F(u,\,v,\,u_1,\,v_1), \, \qquad v_t=-v_{2}+G(u,\,v,\,u_1,\,v_1).
\end{equation}

Besides the well-known NLS equation written as a system of two equations
\begin{equation}\label{NLS}
u_t=-u_{xx}+2 u^2 v, \qquad v_t=v_{xx}-2 v^2 u,
\end{equation}
basic integrable models from a long list of such integrable models are: 
\begin{itemize}[itemsep=0pt]
\item a version of the Boussinesq equation
\[
u_t=u_{2}+(u+v)^{2}, \qquad v_t=-v_{2}+(u+v)^{2};
\]
\item and the two-component form of the Landau-Lifshitz equation
\[
\begin{cases}
\displaystyle u_t=u_{2}-\frac{2 u_1^2}{u+v}-\frac{4\,(p(u,v)\, u_1+r(u)\, v_1)}
{(u+v)^2},
\\[4mm]
\displaystyle v_t=-v_{2}+\frac{2 v_1^2}{u+v}-\frac{4\,(p(u,v)\, v_1+r(-v)\, u_1)}
{(u+v)^2},
\end{cases}
\]
where $r(y)=c_4 y^4+c_3 y^3+c_2 y^2+c_1 y+c_0$ and
\[
p(u,v)=2 c_4 u^2 v^2+c_3 (u v^2-v u^2)-2 c_2 u v+c_1(u-v)+2 c_0.
\]
\end{itemize}
A complete list of integrable systems~\eqref{sys2} up to transformations 
\[
u \to \Phi(u), \qquad v\to \Psi(v)
\]
should contain more the 100 systems. Such a list never has been 
published. In~\cite{MikShaYam87} appears a list complete up to ``almost invertible'' transformations~\cite{MSY}.
All of these equations have a fourth order symmetry of the form 
\begin{equation}
\label{kvazsym}
 \begin{cases}
u_{\tau}=\phantom{-}u_{xxxx}+f(u,v,u_x,v_x,u_{xx},v_{xx},u_{xxx},v_{xxx}), \\[1.5mm]
v_{\tau}=-v_{xxxx}+g(u,v,u_x,v_x,u_{xx},v_{xx},u_{xxx},v_{xxx})
\end{cases}.
\end{equation}
Reference~\cite{SokWol99} contains a classification of integrable equations of the form
\[
 \begin{cases}
u_t = \phantom{-}u_{xx} + A_{1}(u,v)\, u_x + A_{2}(u,v)\, v_x + A_{0}(u,v), \\[1.5mm]
v_t = - v_{xx} + B_{1}(u,v)\, v_x + B_{2}(u,v)\, u_x + B_{0}(u,v) 
 \end{cases}
\]
that includes C-integrable equations. The integrability requirement is that the system admits a fourth order symmetry~\eqref{kvazsym} and triangular systems like~$
u_t=u_{xx}+2 u v_x$, $v_t=-v_{xx}-2 v v_x$ are disregarded.

\subsection{Integrability conditions} 
\label{SectionIntegrabilityConditions}

We denote by $ \mathcal{F} $ a field of functions depending on a finite number of variables $u, u_1, \ldots $. The field of constants is $\C$.

\subsubsection{Pseudo-differential series.}

Consider a skew field of (non-commutative) pseudo-differential series of the form
\begin{equation}\label{serA}
 A=a_{m}D^m+a_{m-1}D^{m-1}+\ldots + a_0+a_{-1}D^{-1}+
 a_{-2}D^{-2}+\ldots\,,
\qquad a_k\in \mathcal{F}\, .
\end{equation}
The number $\ord{A}=m\in \Z$ is called the {\it order} of $A$.  If $a_i=0$ for $i<0$, then $A$ is a~{\it differential operator}.
\medskip 

The product of two pseudo-differential series is defined over monomials by
\[
 D^k\circ b D^m =b D^{m+k}+C_k^1 D(b)D^{k+m-1} + 
 C_k^2 D^2 (b)D^{k+m-2}+
\cdots \, ,
\]
where $k,m\in \mathbb Z$ and $C^j_n=\binom{n}{j}$ is the binomial coefficient. The formally conjugated pseudo-differential series $A^+$ is defined as
\[
 A^+=(-1)^m D^m\circ\, a_{m}+(-1)^{m-1}D^{m-1}\circ\, a_{m-1}+\cdots +
 a_0-D^{-1}\circ\,a_{-1}+D^{-2}\circ\, a_{-2}+\cdots\,.
\]
For any series~\eqref{serA} there is a unique inverse series~$B$
such that $A\circ B=B\circ A=1$, and there are $m$-th roots
$C$ such that $C^m=A$, unique up to a numeric factor~$\varepsilon$  with~$\varepsilon^m=1$. Notice that~$\ord{B}=-m$ and~$\ord{C}=1$.

\begin{definition}
The~\emph{residue} of series~\eqref{serA} is the coefficient of $D^{-1}$:~$\operatorname{res}(A)=a_{-1}$.
The \emph{logarithmic residue} of $A$ is defined as~$
\operatorname{res\, log} A=a_{n-1}/a_n$.
\end{definition}
\begin{theorem}[\hspace{1sp}\cite{adler}] \label{adler}
For any two series $A$, $B$ the
residue of the commutator belongs to $\hbox{Im}\,D$:
\[
{\rm res} [A,B]=D(\sigma (A,B)),
\]
where\quad~$\ds
\sigma (A,B)=\sum_{p\le {\rm ord}(B),\ q\le
{\rm ord}(A)}^{p+q+1>0}C^{p+q+1}_{q}\, \times  
\sum_{s=0}^{p+q}(-1)^s D^s(a_q)D^{p+q-s}(b_q)
$.
\end{theorem}

\subsubsection{Formal symmetry.} 

\begin{definition} A pseudo-differential series~$R$ that satisfies
\begin{equation}\label{Lambdaeq}
D_t(R)=[F_{*},\,R], \qquad \mbox{where} \qquad F_{*}=\sum_{i=0}^{n}
\frac{\partial F}{\partial u_{i}}D^{i}
\end{equation}
is called~\emph{formal symmetry}, or~\emph{formal recursion operator}\footnote{Relation~\eqref{Lambdaeq} can be rewritten as $[D_t-F_{*}^{+}, \, R] = 0$. Therefore any genuine operator that satisfies~\eqref{Lambdaeq}  maps higher symmetries of the equation~\eqref{eveq} to higher symmetries. }, of eq.~\eqref{eveq}.
\end{definition}
\begin{proposition}[\hspace{1sp}\cite{sokshab}]\label{Lambdaspace}\hspace{1sp}
\begin{enumerate}[label=\arabic*),itemsep=1pt]
\item If $R_1$ and $R_2$ are formal symmetries, then~$R_1\circ R_2$ is a formal symmetry too;
\item\label{item:it2p2} If~$R$ is a formal symmetry of order~$k$, so is $R^{i/k}$ for any $i\in \Z$;
\item Let $\Lambda$ be a formal symmetry of order~1
\begin{equation}\label{FSym} \Lambda=l_{1} D+l_{0}+l_{-1}D^{-1}+\cdots .
\end{equation}
 Then $R$ can be written in the form
\[
R=\sum_{-\infty}^k a_i \Lambda^i, \qquad  k=\ord{R}, \quad a_i\in \C; 
\] 
\item In particular, any formal symmetry $\bar \Lambda$ of order~1 has the form
\[
\bar \Lambda = \sum_{-\infty}^1 c_i \Lambda^i, \qquad c_i\in \C.
\]
\end{enumerate}
\end{proposition}
In Sections~\ref{sec:sai} and~\ref{sec:inaeq} we will only  consider a formal symmetry~$\Lambda$ of order~1, without loss of generality  (see Item~\ref{item:it2p2} of Proposition~\ref{Lambdaspace}).
\begin{example}
Consider finding formal symmetries of equations of KdV type
\begin{equation}\label{kdvt}
u_{t}=u_{3}+f(u, u_1).
\end{equation}
With
\[
F_{*}=D^3+\frac{\partial f}{\partial u_{1}} D+
\frac{\partial f}{\partial u},\qquad \quad
\Lambda =l_{1} D+l_{0}+l_{-1}D^{-1}+\cdots.
\]
equation~\eqref{Lambdaeq} becomes an infinite system of differential equations whose members are the coefficients of $D^3, D^2, \dots$ equaled to zero. The first equations are
\begin{gather*}
D^3:\quad 3 D(l_1)=0; \qquad D^2:\quad 3 D^2(l_1)+3 D(l_0)=0; 
\\
D: \quad D^3(l_1)+3 D^2(l_0)+3 D(l_{-1})+
\frac{\partial f}{\partial u_{1}}\, D(l_1)=
 D_t(l_1)+l_1\, D\left(\frac{\partial f}{\partial u_{1}}\right).
\end{gather*}
The first three equations above imply that~$l_1=1$, $l_0=0$ and~$l_{-1}=\frac{1}{3}\, \frac{\partial f}{\partial u_{1}}$, i.e.
\[
\Lambda=D+\frac{1}{3}\,
\frac{\partial f}{\partial u_{1}}D^{-1}+\cdots
\]
The first obstacle for the existence of $\Lambda$ appears in the coefficient of~$D^{-1}$, requiring that~$\partial^4f(u,u_1)/\partial u_1^4=0$. Thus, there are no formal symmetries for an arbitrary function~$f(u,u_1)$ in~\eqref{kdvt}.
\end{example}
\begin{remark}\label{rem24} In general, for any $k$ the equation for the coefficient $l_k$ of $\Lambda$ can be written as $D(l_k)=S_k$, where $S_k\in\mathcal{F}$ is an already known function. The equation is solvable only if $S_k \in {\rm Im}\, D$. Thus there  are infinitely many obstacles for the existence of a formal symmetry. The integration constant appearing in $l_k$ can be taken as~0 (except for~$k=1$) because of Proposition~\ref{Lambdaspace}.
\end{remark}

\begin{theorem}[\hspace{1sp}\cite{ibshab,sokshab}]\label{tlsym}  If an equation $u_t=F$ possesses an infinite sequence of higher symmetries
\[
u_{\tau_{i}}=G_i(u, \dots ,u_{m_{i}}), \qquad m_i \rightarrow \infty
\]
then it has a formal symmetry.
\end{theorem}

\subsubsection{Formal symplectic operator.}\label{ssFor}

It is known~\cite[p.~122]{MikShaSok91} that for any conserved density $\rho$ the variational derivative   
\[
X = \frac{\delta \rho}{\delta u}=\sum_k (-1)^k D^k \left(\frac{\partial \rho}
{\partial u_k}\right) , 
\]
satisfies the equation conjugate to~\eqref{DFsym}:
\begin{equation}\label{varR}
D_t\,(X)+F_*^+\,(X)=0\, .
\end{equation}
Any solution $X \in \mathcal{F}$ of equation~\eqref{varR} is called {\it cosymmetry}.
 
The conjugate concept of a formal symmetry is the following.
\begin{definition} A pseudo-differential series
\[
S=s_m D^m+s_{m-1} D^{m-1}+\cdots +s_0+s_{-1} D^{-1} +\cdots\, ,\qquad s_m\ne 0, \quad s_i\in\mathcal{F}
\]
is called a {\it formal symplectic operator}\footnote{Relation~\eqref{Req} can be rewritten as $(D_t+F_{*}^{+})\circ S=S (D_t-F_{*})$. This means that a genuine operator $S: \mathcal{F}\to \mathcal{F}$ maps symmetries to cosymmetries. If equation~\eqref{eveq} is Hamiltonian, then the symplectic operator, which is inverse to the Hamiltonian operator, satisfies equation~\eqref{Req}~\cite{Dorf}.} of order $m$ for equation~\eqref{eveq}  if it satisfies
\begin{equation}\label{Req}
D_t(S)+S\, F_*+F_*^+\, S=0\, .
\end{equation}
\end{definition}
It follows from~\eqref{Req} that equations of the form~\eqref{eveq} of even order $n$ have no formal symplectic operators.
\begin{lemma}\label{lem:ratsym}
 The ratio
$S_{1}^{-1} S_{2}$ of any two formal symplectic operators $S_1$ and $S_2$ satisfies the equation~\eqref{Lambdaeq} of a formal symmetry.
\end{lemma}

\begin{theorem}[\hspace{1sp}\cite{ibshab,sokshab}]\label{svsok} If equation $u_t=F$ possesses
an infinite sequence of local conservation laws,
then the equation has: 
\begin{enumerate}[label=\arabic*),itemsep=1pt]
\item a formal recursion operator $\Lambda$ and 
\item a formal symplectic operator $S$ of first order.
\end{enumerate}
\end{theorem} 
\begin{remarknn}[see~\cite{sokshab}]
Without loss of generality one can assume that 
\[
 S^{+} = -S,  \qquad  \Lambda^{+} = -S^{-1} \Lambda S.
\]
\end{remarknn}

\subsubsection{Canonical densities and necessary integrability conditions.}

In this paragraph we formulate the necessary conditions over equations~\eqref{eveq} to admit infinite higher symmetries or conservation laws. According to Theorems~\ref{tlsym} and \ref{svsok}, such equations possess a formal symmetry. The obstructions in Remark~\ref{rem24} to the existence of a formal symmetry, are thus integrability conditions. It tuns out that these conditions can be written in the form of conservation laws. 

\begin{definition}\label{drho} For equations~\eqref{eveq} possessing a formal symmetry~$\Lambda$, the functions
\begin{equation}\label{rro}
\rho_i={\rm res}\,(\Lambda^i), \qquad i=-1,1,2,\dots, \, \quad {\rm and}
\qquad \rho_0={\rm res}\, \log (\Lambda)
\end{equation}
are called {\sl canonical densities} for equation~\eqref{eveq} .
\end{definition}
Adler's theorem \ref{adler} implies the following result.
\begin{theorem}\label{prho} If an equation~\eqref{eveq} has a formal
symmetry $\Lambda$, then the canonical densities~\eqref{rro}
define corresponding local conservation laws
\begin{equation}\label{canlaws}
D_{t}(\rho_i)=D(\sigma_{i}), \quad \sigma_{i} \in \mathcal{F}, \qquad
i=-1,0,1,2,\dots.
\end{equation}
\end{theorem}

\begin{theorem}[\hspace{1sp}\cite{sokshab}]\label{contriv}
Under the assumptions of Theorem \ref{svsok}, all even canonical densities $\rho_{2j}$ belong to ${\rm Im}\, D$.
\end{theorem}

\begin{example} The differential operator $\Lambda = D$ is a formal symmetry for any linear equation of the form $u_t=u_n$. Therefore 
all canonical densities are equal to zero.
\end{example}
\begin{example} \label{kdvl} The KdV equation~\eqref{kdv} 
has a recursion operator 
\begin{equation}\label{recop}
\hat{\Lambda}=D^2+4 u+2 u_1 D^{-1}\, ,
\end{equation}
which satisfies equation~\eqref{Lambdaeq}.
A corresponding formal symmetry of order~1 for the KdV equation is $\Lambda =\hat{\Lambda}^{1/2}$. The infinite commutative hierarchy of  symmetries for the KdV equation is generated by the recursion operator:
\[
G_{2k+1}=\hat{\Lambda}^k(u_1)\, .
\]
The first five canonical densities for the KdV equation are
\[
\rho_{-1}=1,\qquad\rho_0=0,\qquad \rho_1=2 u, \qquad \rho_2=2 u_1, \qquad
\rho_3=2 u_2+ u^2.
\]
%We see that even canonical densities are trivial. 
\end{example}

\begin{example}
The Burgers equation~\eqref{burgers} has the recursion operator
\[ \Lambda=D+u+u_1 D^{-1}\, .\]
Functions $G_n=\Lambda^n(u_1)$ are generators of symmetries for the
Burgers equation. The canonical densities for the Burgers equation are
\[
\rho_{-1}=1,\qquad\rho_0=u,\qquad \rho_1=u_{1}, \qquad
\rho_2=u_2+ 2 u u_{1}, \dots\, .
\]
 Although $\rho_{0}$ is not trivial (i.e.~$\rho_0\notin\operatorname{Im}{D}$), all other canonical densities are trivial.
\end{example}

Now we can refine the definition of $C$-integrability such that linear equations become $C$-integrable. 
\begin{definition}\label{clar}
Equation~\eqref{eveq} is called {\it $S$-integrable}   if it has a formal symmetry that provides infinitely many non-trivial canonical densities.  An equation is called {\it $C$-integrable}  if it has a formal symmetry such that only finite number of canonical densities are non-trivial.
\end{definition}
\begin{remarknn}
It follows from Theorems \ref{tlsym}, \ref{svsok}, \ref{prho} and \ref{contriv} that if we are going  to find equations~\eqref{eveq} with higher symmetries, we have to use conditions~\eqref{canlaws} only, while for equations with higher conservation laws we may additionally assume that $\rho_{2 j}=D(\theta_{j})+c_j,$ where $\theta_{j}\in \mathcal{F}$ and $c_j \in\C$. 
Thus the necessary conditions, which we employ for conservation laws are stronger than ones for symmetries.  
\end{remarknn}

Using the ideas of~\cite{CheLeeLiu79, mesh}, a recursive formula for the whole  infinite chain of canonical conserved densities  can be derived.  For equations of the form~\eqref{vvs} such a formula was obtained in~\cite{meshsok}:
\begin{align}
\rho_0&=-\frac{1}{3}f_{2},\qquad \rho_1=\frac{1}{9}f_{2}^2-\frac{1}{3}f_{1}+\frac{1}{3} D(f_{2}) ,\nonumber\\
\rho_{n+2}&=\frac{1}{3}\bigg[\sigma_n-\delta_{n,0}f_{0} -f_{1}\rho_{n} - f_{2}\Big(D(\rho_{n}) + 2\rho_{n+1}+\sum_{s=0}^{n} \rho_{s}\,\rho_{n-s}\Big)\bigg]
\nonumber\\
&-\sum_{s=0}^{n+1} \rho_{s}\,\rho_{n+1-s}
-\frac{1}{3}\sum_{0\le s+k\le n}\rho_{s}\,\rho_{k}\,\rho_{n-s-k}
\nonumber\\
&-D\biggl[\rho_{n+1}+\frac{1}{2}\sum_{s=0}^{n}\rho_{s}\,\rho_{n-s}+
\frac{1}{3} D(\rho_{n}) \biggr], \qquad n\ge0, \label{rekkur_sc}
\end{align}
Here, $\delta_{i,j}$ is the Kronecker delta and~$f_{i}=\p f\!/\p u_i$, for $i=0,1,2$. 

Using the integrability conditions, one can find all equations of the prescribed type, which have a formal symmetry.  A full classification result includes:
\begin{enumerate}[label=\arabic*),itemsep=1pt]
\item A complete\footnote{usually complete up to a class of admissible transformations.} list of integrable equations that satisfy the necessary integrability conditions;
\item\label{it:confint} A confirmation of integrability for each equation from the list;
\end{enumerate}
For Item~\ref{it:confint} one can find a Lax representation or a transformation that links the equation with an equation known to be integrable. The existence of an auto-B\"acklud transformation with an arbitrary parameter is also a proper justification of integrability.

\begin{remarknn}  From a proof of a classification result one can derive a constructive description of transformations that bring a given integrable equation to one from the list and the number of necessary conditions, which should be verified for a given equation to establish its integrability. 
\end{remarknn}

The classifications of integrable evolution equations discussed in Section~\ref{subs:someclass} have been performed using the theory described in this section.

\subsection{Recursion and Hamiltonian quasi-local operators.}\label{recHam}
 
Recursion and Hamiltonian operators establish additional relations between higher symmetries and conserved densities. 

\begin{proposition}\label{propdefrec} If the operator $\mathcal{R}: \mathcal{F}\to \mathcal{F}$ satisfies the equation\footnote{In the language of differential geometry, this relation means that the Lie derivative   of the  operator $\mathcal{R} $, by virtue of equation~\eqref{eveq}, is zero.}
\begin{equation}\label{defR}
D_t(\mathcal{R})=F_*\, \mathcal{R}-\mathcal{R}\,F_*,
\end{equation}
then, for any symmetry\footnote{For brevity, we often refer to the symmetry by its generator~$G$.}~$G$ of equation~\eqref{eveq}, $\mathcal{R}(G)$ is also a symmetry of~\eqref{eveq}.
\end{proposition}
\begin{definition} An operator $\mathcal{R}: \mathcal{F}\to \mathcal{F}$ satisfying~\eqref{defR} is called {\it recursion operator} for equation~\eqref{eveq}.
\end{definition}
The set of all recursion operators forms an associative algebra over~$\C$.

The simplest symmetry for any equation~\eqref{eveq}
is $u_{\tau}=u_x$. Acting with a recursion operator over~$u_x$ usually yields the generators of all the other symmetries. 

A recursion operator is usually non-local (see, for instance, 
\eqref{recop}) so it can only be applied to a very special subset of~$\mathcal F$ to get a function in~$\mathcal F$.
 
\subsubsection{Quasi-local recursion operators.}

Most of the known recursion operators have the following special non-local structure:
\begin{equation}\label{anz}
\mathcal{R}=R+\sum_{i=1}^k G_i\, D^{-1}\circ g_i,\qquad g_i, G_i\in \mathcal{F}, 
\end{equation}
where $R$ is a differential operator. Such operators are called \emph{quasi-local} or \emph{weakly nonlocal}.

\begin{definition} \label{def214} An operator $\mathcal{R}$ of the form~\eqref{anz} is called a {\it quasi-local recursion operator} for equation~\eqref{eveq} if
\begin{enumerate}[label=\arabic*),itemsep=1pt]
\item\label{item:qsro1}  $\mathcal{R}$, considered as a pseudo-differential series, satisfies~\eqref{defR};
\item  the functions $G_i$ are generators of some symmetries for~\eqref{eveq};
\item\label{item:qsrof}  the functions $g_i$ are variational derivatives of conserved
densities\footnote{It might be reasonable to add  the hereditary property {\rm~\cite{Fuch}} of the operator $\mathcal{R}$ to the properties \ref{item:qsro1}--\ref{item:qsrof}.}.
\end{enumerate}
\end{definition}
\begin{example}
The recursion operator~\eqref{recop} for the KdV equation is quasi-local with $k = 1,\, G_1 = {u_x}/{2}$ and $g_1 ={\delta u}/{\delta u}= 1$.
\end{example}
The first reference we know of where a quasi-local ansatz for finding a recursion operator was used, is~\cite{sokkn}. 

It can be proved that the set of all quasi-local recursion operators for the KdV equation form a commutative associative algebra~$A_{\rm rec}$ over $\C$ generated by the operator~\eqref{recop}, i.e.~$A_{\rm rec}$  is isomorphic to the algebra of all polynomials in one variable.

It turns out that this is not true for integrable models such as the Krichever-Novikov and the Landau-Lifshitz equations. In particular, the  Krichever-Novikov equation~\eqref{KN} has two quasi-local recursion operators
$\mathcal{R}_1$ and $\mathcal{R}_2$ such that 
$
\mathcal{R}_2^2=\mathcal{R}_1^3-\phi \mathcal{R}_1-\theta,  
$
where the constants $\phi, \theta$ are polynomial in the coefficients of $Q$.

\begin{remarknn} In the case of the the  Krichever-Novikov equation~\eqref{KN}, the ratio $\mathcal{R}_3=\mathcal{R}_2 {\mathcal R}_1^{-1}$ satisfies equation~\eqref{defR}. It belongs to the
skew field of differential operator fractions~\cite{ore}.
However, this operator is not quasi-local and it is unclear how to
apply it even to the simplest symmetry generator $u_x$.
\end{remarknn}

Further information about this matter can be found in~\cite{demsok,sokkn}.

\subsubsection{Hamiltonian operators.} 

Most of the known integrable equations~\eqref{eveq} can be written in a Hamiltonian form as
\[
u_t=\mathcal{H}\left(\frac{\delta \rho}{\delta u}  \right),
\]
where $\rho$ is a conserved density and $\mathcal{H}$ is a Hamiltonian
operator.  The analog of the operator identity~\eqref{defR} for Hamiltonian operators is given by
\begin{equation}\label{Heq}
(D_t-F_*) \, \mathcal{H}= \mathcal{H} (D_{t}+F_{*}^+),
\end{equation}
which means that $\mathcal H$ maps cosymmetries to symmetries. This formula justified by the general theory~\cite{Dorf}, that states that Hamiltonian operators play an inverse role of that of symplectic operators (see~\eqref{Req}).

The Poisson bracket  corresponding to a Hamiltonian operator $\mathcal{H}$ is defined by 
\begin{equation}\label{PDEbrop}
\{f,\,g\}= \frac{\delta f}{\delta u}  \, \mathcal{H} \Big(\frac{\delta g}{\delta u} \Big).
\end{equation}
Skew-symmetricity and the Jacobi identity for~\eqref{PDEbrop} are required. Namely, 
\be \label{HHcond1}
\{f,g\}+\{g,f\} \in {\rm Im}\,D , 
\ee
\be \label{HHcond2}
\{\{f,g\},h\}+\{\{g,h\},f\}+\{\{h,f\},g\} \in {\rm Im}\,D
\ee
for $f,g\in\mathcal{F}$.
Thus,  Hamiltonian operators should satisfy,  besides~\eqref{Heq}, some identities (see for example~\cite{Dorf,Olv93}) equivalent to~\eqref{HHcond1},~\eqref{HHcond2}. Lemma~\ref{lem:ratsym} suggests the next result.
\begin{lemma} If operators $\mathcal{H}_1$ and $\mathcal{H}_2$ satisfy~\eqref{Heq}, then $\mathcal{R}=\mathcal{H}_2
\mathcal{H}_1^{-1}$ satisfies~\eqref{Lambdaeq}.
\end{lemma}

As a rule, Hamiltonian operators are local (differential)
or quasi-local operators
\[\mathcal{H}=H+\sum_{i=1}^m G_i D^{-1} \bar G_i,
\]
where $H$ is a differential operator and $G_i, \bar G_i$ are  
symmetries~\cite{mokfer, malnov}.

The KdV equation possesses two local Hamiltonian operators  
\[
\mathcal{H}_1=D, \qquad \mathcal{H}_2=D^3+4 u D+2 u_x.
\]
Their ratio gives the recursion operator~\eqref{recop}.
   
The first example 
\[
\mathcal{H}_0=u_x D^{-1} u_x
\]
of  a quasi-local Hamiltonian operator  was found in paper~\cite{sokkn}, where the Krichever-Novikov equation~\eqref{KN} was studied. In~\cite{demsok} it was shown that, besides $\mathcal{H}_0$,   equation~\eqref{KN} possesses two more quasi-local Hamiltonian operators.

\section{Integrable non-abelian equations}\label{sec:inaeq}

\subsection{ODEs on free associative algebras}

We consider ODE systems of the form
\begin{equation}\label{geneq}
\frac{d x_{\alpha}}{d t}=F_{\alpha}({\bf x}), \qquad {\bf x}=(x_1,...,x_N), \qquad \alpha=1,\ldots,N,
\end{equation}
where $x_i(t)$ are $m\times m$ matrices, $F_{\alpha}$ are (non-commutative) polynomials with constant scalar coefficients. As usual, a symmetry is defined as an equation
\begin{equation}\label{gensymmat}
\frac{d x_{\alpha}}{d \tau}=G_{\alpha}({\bf x}), 
\end{equation}
compatible with~\eqref{geneq}. 

In the case $N=2$ we denote $\,x_1=u,\, x_2=v$.
\subsubsection{Manakov top.}

The  system 
\begin{equation}\label{man}
u_{t}=u^2 \, v-v \, u^2, \qquad v_{t}=0
\end{equation}
has infinitely many symmetries for any size of the matrices $u$ and $v$. Many important {\it multi-component} integrable systems
  can be obtained as
reductions of~\eqref{man}. 
For instance, if $u$ is $m\times m$ matrix such that $u^t=-u$, and $v$ is a constant diagonal matrix, then 
\eqref{man} is equivalent to the $m$-dimensional Euler top. The
integrability of this model by the inverse scattering method was established by S.V. Manakov in~\cite{man}. 

Consider the cyclic reduction
\[
u=\left( \begin{array}{cccccc}
0&u_1&0&0&\cdot&0\\
0&0&u_2&0&\cdot&0\\
\cdot&\cdot&\cdot&\cdot&\cdot&\cdot\\
0&0&0&0&\cdot&u_{m-1}\\
u_{m}&0&0&0&\cdot&0
\end{array}\right)\, , \qquad 
v=\left( \begin{array}{cccccc}
0&0&0&\cdot&0&J_m\\
J_1&0&0&\cdot&0&0\\
0&J_2&0&\cdot&0&0\\
\cdot&\cdot&\cdot&\cdot&\cdot&\cdot\\
0&0&0&\cdot&J_{m-1}&0
\end{array}\right)\, ,
\]
where $u_k$ and $J_k$ are matrices of lower size.
Then~\eqref{man} is equivalent to   the non-abelian Volterra chain
\[
\frac{d}{dt}u_k=u_k  u_{k+1}  J_{k+1}-J_{k-1}  u_{k-1}  u_k, \qquad k=1,\ldots,m.
\]
If we assume $m=3,\, J_1=J_2=J_3= \,{\rm Id}$ and
$u_3=-u_1 -u_2$ the system becomes
\[
u_t=u^2+u  v+v  u\, ,\qquad v_t=-v^2-u  v-v  u\, .
\]

\subsubsection{Matrix generalization of a flow on an elliptic curve.}
The system 
\begin{equation}\label{inuv}
\left\{
\begin{aligned}
u_t  &=  v^2 + c u + a \, {\rm I},
\\
v_t  &=  u^2 - c v + b \, {\rm I},
\end{aligned}  \qquad   a , b, c \in \C,
\right.
\end{equation}
where $u$ and $v$  are $m\times m$-matrices and~${\rm I}$ is the identity matrix, is integrable for any~$m$. It has a Lax pair~\cite{sokwol3} and possesses an infinite sequence of polynomial symmetries. If $m=1$, this system can be written in the Hamiltonian form 
\[
u_t = - \frac{\partial H}{\partial v}, \qquad v_t =  \frac{\partial H}{\partial u}
\]
with Hamiltonian  
\[
H = \frac{1}{3} u^3 - \frac{1}{3} v^3 - c u v + b u - a v.
\]
For generic $a,b,c$ the relation $H = \text{constant}$ defines an elliptic curve, and equations~\eqref{inuv} describe the motion of a point along this curve. 

In the homogeneous case the system~\eqref{inuv} has the form~\cite{miksokcmp}
\begin{equation}\label{uv}
u_{t}=v^2, \qquad v_{t}=u^2.
\end{equation}
F.~Calogero has observed that  in the matrix case the functions $x_i=\lambda_i^{1/2}$, where~$\lambda_i$ are the eigenvalues of the matrix  $u-v,$ satisfy  the following integrable system:
\[
x_i^{''}=-x_i^5+\sum_{j\ne i} \Big[(x_i-x_j)^{-3}+(x_i+x_j)^{-3}\Big].
\] 

\subsubsection{Non-abelian systems.}
 The variables $x_1,\dots,x_N$ in~\eqref{geneq} can be regarded as generators of a free associative algebra $\mathcal{A}$. We call systems on~$\mathcal{A}$ {\it non-abelian systems}. 
In order to understand what compatibility of equations~\eqref{geneq} and~\eqref{gensymmat} means, we use the following definition.
\begin{definition}
A linear map $d: \mathcal{A}\to \mathcal{A}$ is called a derivation if it satisfies the 
Leibnitz rule: $d (x y)=x d(y)+d(x) y$.
\end{definition}
Fixing~$d(x_i)=F_i({\bf x})$ over all generators~$x_i$ of~$\mathcal{A}$ uniquely determines~$d(z)$ for any~$z\in \mathcal{A}$, through the Leibnitz rule. The polynomials $F_i$ can be taken arbitrarily. 
Instead of dynamical system~\eqref{geneq} one considers the derivation $D_t: \mathcal{A}\to \mathcal{A}$ such that $D_t(x_i)=F_i$. Compatibility of~\eqref{geneq} and~\eqref{gensymmat} means that the corresponding derivations $D_t$ and $D_{\tau}$ commute: $D_t D_{\tau}-D_{\tau} D_t=0$.

From the symmetry approach point of view, system~\eqref{geneq} is integrable if it possesses infinitely many linearly independent symmetries.

\paragraph{Two-component non-abelian systems.}

Consider non-abelian systems 
\[
u_t=P(u,v)\, ,\qquad v_t=Q(u,v)\, ,\qquad P,Q\in \mathcal{A}
\]
on the free associative algebra $\mathcal{A}$ over $\C$ with generators $u$ and $v$. 
Define an involution
$\star$ on $\mathcal{A}$ by the formulas
\begin{equation}\label{star1}
u^\star=u\, ,\quad v^\star=v\, ,\quad (a\, b)^\star=
b^\star \, a^\star \, ,\quad a,b\in \mathcal{A}.
\end{equation}
Two systems related to each other by a linear
transformation of the form
\begin{equation}\label{invert}
\hat{u}=\alpha u+\beta v\, ,\qquad \hat{v}=\gamma u+\delta v\, ,\qquad
\alpha \delta-\beta\gamma \ne 0
\end{equation}
and involutions~\eqref{star1} are defined as {\it equivalent}.

The simplest class is that of quadratic systems of the form
\begin{equation} \label{equ}
\left\{
\begin{aligned}
u_t   &=  \alpha_1 u \, u  + \alpha_2 u \, v + \alpha_3  v \, u +
\alpha_4 v \, v,\\
v_t   &=  \beta_1 v  \, v  + \beta_2 v  \, u + \beta_3  u \, v + \beta_4
u \, u.
\end{aligned}
\right.
\end{equation}
The problem is to describe all non-equivalent systems~\eqref{equ} which possess infinitely many symmetries.
Some preliminary results were obtained in~\cite{miksokcmp}. Here we follow the paper~\cite{sokwol3}.

It is reasonable to assume that the corresponding scalar system 
\begin{equation}\label{absys}
\left\{
\begin{aligned}
u_t &= a_1 u^2 + a_2 u v + a_3 v^2, \\
v_t &= b_1 v^2 + b_2  u v + b_3 u^2 ,
\end{aligned}
\right.
\end{equation}
where
$a_1=\alpha_1$,\ \ $a_2=\alpha_2+\alpha_3$,\ \ $a_3=\alpha_4$,\ \ $b_1=\beta_1$,\ \ $b_2=\beta_2+\beta_3$,\ \ $b_3=\beta_4$ should be integrable. 
The main feature of integrable systems of the form~\eqref{absys} is the existence of infinitesimal polynomial symmetries and first integrals. Another evidence of integrability is the absence of movable singularities in solutions for complex~$t$. The so-called Painlev\'e approach is based on this assumption.  
Our first requirement is that system~\eqref{absys} should possess a polynomial first integral $I$. 
\begin{lemma} Suppose that a system~\eqref{absys} has a homogeneous polynomial integral~$I(u,v)$.
Then it has an infinite sequence of polynomial symmetries of the form
\begin{equation}\label{infsym}
\left\{
\begin{aligned}
u_{\tau} &= I^N\,(\alpha_1 u^2 + \alpha_2 u v + \alpha_3 v^2), 
\\
v_{\tau} &= I^N\, (\beta_1 v^2 + \beta_2  u v + \beta_3 u^2)
\end{aligned}
\right. \qquad N\in \N.
\end{equation}
\end{lemma}
Writing~$I$ in factorized form:
\[
I = \prod_{i=1}^k (u-\kappa_i v)^{n_{i}}, \qquad n_i \in \N,  \qquad \kappa_i\ne \kappa_j \,\, {\rm if} \,\,i\ne j.
\]

\begin{theorem} \label{prop4} Suppose that at least one of the coefficients of a system~\eqref{absys} is not equal to zero. Then $k\le 3. $ 
\end{theorem}
Consider the case $k=3$.\footnote{For the  cases $k=1,2$ see~\cite{sokwol3}.} Transformations~\eqref{invert} reduce $I$ to the form
\begin{equation} \label{3roots}
I=u^{k_1}(u-v)^{k_2} v^{k_3},
\end{equation} 
where $k_i$ are natural numbers which are defined up to permutations. Without loss of generality we assume that 
\[k_1\le k_2\le k_3
\]
and that $k_1,k_2,k_3$ have no non-trivial common divisor. 
\begin{lemma}  A system~\eqref{absys} has an integral~\eqref{3roots} iff up to a scaling $u\to \mu u,\, v \to \mu v$ it has the following form:
\begin{equation}\label{sys3root}
\left\{
\begin{aligned}
u_t &=  -k_3 \, u^2 + (k_3+k_2) \,u v  \\
v_t &= - k_1 \, v^2 + (k_1+k_2) \, u v.
\end{aligned}
\right.
\end{equation}
\end{lemma}
\begin{proposition} A system~\eqref{sys3root} satisfies the Painlev\'e test in the following three cases:
\begin{enumerate}[label=\text{Case }\arabic*.,ref=\text{Case }\arabic*,leftmargin=2cm,itemsep=1pt]
\item\label{item:case1} $k_1=k_2=k_3=1$;
\item\label{item:case2} $k_1=k_3=1, \quad k_2=2$;
\item\label{item:case3} $k_1=1,\quad k_2=2,\quad k_3=3$.
\end{enumerate}
\end{proposition}
Any non-abelian system which coincides with~\eqref{sys3root} in the scalar case, has the form
\begin{equation}\label{3rootN}
\left\{
\begin{aligned}
u_t &=  -k_3 \, u^2 + (k_2+k_3) \,u v  + \alpha (u v - v u)\\
v_t &= - k_1 \, v^2 + (k_1+k_2) \, v u + \beta (v u - u v).
\end{aligned}
\right.
\end{equation}
Let us find the parameters $\alpha$ and $\beta$ such that~\eqref{3rootN} has infinitely many symmetries that reduce to~\eqref{infsym} in the scalar case.

Consider~\ref{item:case1}:  $k_1=k_2=k_3=1$. The integral $I$ is of degree 3 and~\eqref{infsym} implies that the simplest symmetry is supposed to be of fifth degree.
\begin{theorem} \label{theo3}  In the case $k_1=k_2=k_3=1$ there exist only 5 non-equivalent non-abelian systems of the form 
\eqref{3rootN} that have a fifth degree symmetry. They correspond to the following pairs $\alpha, \beta$ in~\eqref{3rootN}:
\begin{enumerate}[label=\arabic*.,ref=\arabic*,itemsep=1pt]
\item\label{item:th9}\quad $\alpha = -1$,\quad $\beta = -1$,
\item\quad $\alpha = \phantom{-}0$,\quad $\beta = -1$,
\item\quad $\alpha = \phantom{-}0$,\quad $\beta  = - 2$,
\item\quad $\alpha = \phantom{-}0$,\quad $\beta = \phantom{-}0$,
\item\quad $\alpha = \phantom{-}0$,\quad $\beta = -3$.
\end{enumerate}
\end{theorem}
System~\eqref{uv} is equivalent to the system in Item~\ref{item:th9}.

Consider now~\ref{item:case2}: $k_1=k_3=1, \, k_2=2$. We thus suppose a simplest symmetry of degree 6.

\begin{theorem} \label{theo4}  In the case $k_1=k_3=1, \, k_2=2$ there exist only 4  non-equivalent systems~\eqref{3rootN} that have the symmetry of degree six. They correspond to:
\begin{enumerate}[label=\arabic*.,itemsep=1pt]
\item\quad $\alpha = -1$, \quad  $\beta = -1$,
\item\quad $\alpha = \phantom{-}0$, \quad  $\beta = -2$,
\item\quad $\alpha = \phantom{-}0$, \quad  $\beta = \phantom{-}0$,
\item\quad $\alpha = \phantom{-}0$, \quad  $\beta = -4$.
\end{enumerate}
\end{theorem}
The classification of integrable non-abelian systems~\eqref{3rootN} ends with~\ref{item:case3}:
\begin{theorem} \label{theo5}  In the case  $k_1=1,\, k_2=2, \, k_3=3$ there exist only 5  non-equivalent systems~\eqref{3rootN} with the symmetry of degree 8. They correspond to:
\begin{enumerate}[label=\arabic*.,itemsep=1pt]
\item\quad $\alpha = -2$, \quad  $\beta = \phantom{-}0$,
\item\quad $\alpha = -4$, \quad  $\beta = \phantom{-}0$,
\item\quad $\alpha = -6$, \quad  $\beta = \phantom{-}0$,
\item\quad $\alpha = \phantom{-}0$, \quad $\beta = -6$,
\item\quad $\alpha = \phantom{-}0$, \quad $\beta = \phantom{-}0$.
\end{enumerate}
\end{theorem}
The integrable systems found in~\cite{sokwol3} contain all examples from~\cite{miksokcmp} as well as new integrable non-abelian systems of the form~\eqref{equ}. Moreover, all integrable inhomogeneous generalizations of these systems were found in~\cite{sokwol3}. System~\eqref{inuv} is one of them. 

There are interesting integrable non-abelian  Laurent systems. In this case we  extend the free associative algebra  ${\mathfrak A}$ with generators $u$ and $v$ by new symbols $u^{-1}$ and $v^{-1}$ such that
$u u^{-1} = u^{-1} u = v v^{-1} = v^{-1} v = {\rm I}$.  

\begin{example}\label{rem13} In the paper~\cite{WoEf} the following integrable non-abelian  Laurent system 
\[
u_t = u v - u v^{-1} - v^{-1}, \qquad v_t = - v u + v u^{-1} + u^{-1},
\]
proposed by M. Kontsevich, was investigated. It can be regarded as a non-trivial deformation of the integrable system\footnote{In the scalar case this system has a first integral of first degree.} 
\[
u_t = u v, \qquad v_t = - v u 
\]
by Laurent terms of smaller degree.  
\end{example}

\subsection{PDEs on free associative algebra}

In this subsection we consider the so called non-abelian evolution equations, which are natural generalizations of evolution matrix equations.  

\subsubsection{Matrix integrable equations.}
The matrix KdV equation has the following form
\begin{equation}\label{matkdv}
{\bf U}_{t}={\bf U}_{xxx}+3\, ({\bf U} {\bf U}_{x}+{\bf U}_{x} {\bf U}),
\end{equation}
where ${\bf U}(x,t)$ is unknown $m\times m$-matrix. It is known that  this equation has infinitely many 
higher symmetries for arbitrary $m$. All of them can be written in
matrix form.
The simplest higher symmetry of~\eqref{matkdv} is given by
\begin{multline*}
{\bf U}_{\tau}={\bf U}_{xxxxx}+5\, ({\bf U} {\bf U}_{xxx}+{\bf U}_{xxx} {\bf U})+10\,({\bf U}_{x} {\bf U}_{xx}+{\bf U}_{xx} {\bf U}_{x})
\\[2mm]
+10\,({\bf U}^{2} {\bf U}_{x}+{\bf U} {\bf U}_{x} {\bf U}+{\bf U}_{x}{\bf U}^{2}).
\end{multline*}
For $m=1$ this matrix hierarchy of symmetries coincides with the usual KdV hierarchy.

In general~\cite{OlvSok98}, we may consider matrix equations of the form
\[
{\bf U}_{t}=F({\bf U},\, {\bf U}_{1},\, \dots, \,{\bf U}_{n}), \qquad {\bf U}_i=\frac{\partial^i {\bf U}}{\partial x^i},
\]
where $F$ is a (non-commutative) polynomial with constant scalar coefficients. The criterion of integrability is the existence of matrix  higher symmetries 
\[
{\bf U}_{\tau}=G({\bf U},\, {\bf U}_{1},\, \dots, \,{\bf U}_{m}).
\]

The matrix KdV equation is not an isolated example. Many known integrable models have matrix generalizations~\cite{kuper,march,OlvSok98}. In particular, the mKdV equation~$u_t=u_{xxx}+u^2 u_x$ has two different matrix generalizations:
\[
{\bf U}_t={\bf U}_{xxx}+3 {\bf U}^2 {\bf U}_x+3 {\bf U}_x {\bf U}^2,
\]
and (see~\cite{kuper})
\[{\bf U}_t={\bf U}_{xxx}+3 [{\bf U}, {\bf U}_{xx}]-6 {\bf U} {\bf U}_x {\bf U}.
\]
The matrix generalization of the NLS equation~\eqref{NLS} is given by
\[
{\bf U}_t={\bf U}_{xx}-2\, {\bf U} {\bf V} {\bf U}, \qquad {\bf V}_t=-{\bf V}_{xx}+2\, {\bf V} {\bf U} {\bf V}.
\]
The Krichever-Novikov equation~\eqref{KN} with $Q=0$ is called the {\it Schwartz KdV equation}. Its matrix generalization is given by
\[
{\bf U}_t={\bf U}_{xxx}-\frac32\, {\bf U}_{xx} {\bf U}_x^{-1} {\bf U}_{xx}.
\]
The Krichever-Novikov equation with the generic $Q$  has probably no matrix generalizations. 

The matrix Heisenberg equation has the form
\[
{\bf U}_t = {\bf U}_{xx} -2\, {\bf U}_x ({\bf U}+{\bf V})^{-1} {\bf U}_{x},
\qquad {\bf V}_t = -{\bf V}_{xx} +2\, {\bf V}_x ({\bf U}+{\bf V})^{-1} {\bf V}_{x}. 
\]

One of the most renowned hyperbolic matrix integrable equations is the principal chiral $\sigma$-model
\[
{\bf U}_{xy}=\frac12\, ({\bf U}_x {\bf U}^{-1} {\bf U}_y+ {\bf U}_y {\bf U}^{-1} {\bf U}_x).
\]

The system
\[
\begin{aligned}
{\bf U}_t &= \lambda_1 {\bf U}_{x} + (\lambda_2-\lambda_3) {\bf W}^t  {\bf V}^t , \\[2mm]
{\bf V}_t &= \lambda_2 {\bf V}_{x} + (\lambda_3-\lambda_1) {\bf U}^t  {\bf W}^t , \\[2mm]
{\bf W}_t &=\lambda_3 {\bf W}_{x} + (\lambda_1-\lambda_2) {\bf V}^t  {\bf U}^t
\end{aligned}
\]
is a matrix generalization of the 3-wave model. In contrast with the previous equations, it contains matrix transpositions, denoted by $^t$.

Let ${\bf e}_1,\dots, {\bf e}_N$ be a basis of some associative algebra $\mathcal{B}$ and  
\begin{equation}\label{UU}
U=\sum^N_{i=1} u_i \, {\bf e}_i.
\end{equation}
Then, all the matrix equations presented above give rise to corresponding integrable systems in the unknown functions $u_1,
\dots, u_N$ in~\eqref{UU}. Indeed, just the associativity of the product in~$\mathcal B$ is enough to ensure that the symmetries of a matrix equation remain being symmetries of the corresponding system for $u_i$.

In the matrix case $\mathcal{B}={\mathfrak gl}_m$ interesting examples of integrable multi-component systems are produced by Clifford algebras and by group algebras of associative rings. 

The most fundamental setting for the non-abelian equations is the formalism of free associative algebras, leading to a generalization of matrix equations.

\subsubsection{Non-abelian evolution equations over free associative algebras.}\label{naev}

Let us consider evolution equations on an infinitely generated free associative algebra~$\mathcal{A}$. In the case of
one-field non-abelian equations the generators of $\mathcal{A}$ are denoted by
\begin{equation}\label{generat}
U, \quad U_{1}=U_{x}, \quad \dots, \quad U_{k}, \quad \dots .
\end{equation}
Being $\mathcal A$ free, no algebraic relations between the generators exist. All definitions can be easily generalized to the case of several non-abelian variables.

The formula 
\begin{equation}\label{nonabel}
U_{t}=F(U,\, U_{1},\, \dots, \,U_{n}), \qquad F\in \mathcal{A} 
\end{equation} 
defines a derivation $D_t$ of  $\mathcal{A}$ which commutes with the basic derivation
\[
D=\sum_{0}^{\infty} U_{i+1} \frac{\partial}{\partial U_{i}}.
\]
It is easy to check that $D_t$ is defined by the vector field
\[
D_t= 
\sum_0^{\infty} D^{i} (F) \frac{\partial}{\partial U_i}.
\]
The concepts of symmetry, conservation law, the operation $^*$, and formal symmetry have to be specified for differential equations on free associative algebras. 

As in the scalar case, a symmetry is an evolution equation
\[
U_{\tau}=G(U,\, U_{1},\, \dots, \,U_{m}),
\]
such that the vector field
\[
D_{G}=\sum_0^{\infty} D^{i} (G) \frac{\partial}{\partial u_i}
\]
commutes with $D_t$. The polynomial $G$ is called {\it the symmetry generator}.

The condition $[D_t, D_{G}]=0$ is equivalent to $D_t(G)=D_{G}(F)$. The latter
relation can be rewritten as
\[
G_{*}(F)-F_{*}(G)=0,
\]
where the differential operator $H_{*}$ for any  $H\in \mathcal{A}$ can be defined as follows.

For any $a \in \mathcal{A}$ we denote by $L_a$ and $R_a$ the operators of left and right multiplication by $a$:
\[
L_a(X)=a\,X, \qquad R_a(X)=X\,a, \qquad X\in \mathcal{A}.
\]
The associativity of $\mathcal{A}$ is equivalent to the identity $[L_a, R_b]=0$
for any $a$ and $b$. Moreover,
\[
L_{ab}=L_{a}\,L_{b}, \qquad R_{ab}=R_{b}\,R_{a},
\qquad L_{a+b}=L_{a}+L_{b},   \qquad R_{a+b}=R_{a}+R_{b}.
\]
\begin{definition}\label{localr}
We denote by $\mathcal{O}$ the associative algebra generated by
all operators of left and right multiplication by any element~\eqref{generat}.
This algebra is called the {\it algebra of local operators}.
\end{definition}

Extending the set of generators with an additional non-commutative symbol~$V_0$ and prolonged symbols $V_{i+1}=D(V_i)$, one can define, 
given~$H(U, U_1, \dots, U_k) \in \mathcal{A}$,
\[
H_{*}(V_0)=\frac{\partial}{\partial \varepsilon} H(U+\varepsilon V_0, \
U_1+\varepsilon V_1, \
U_2+\varepsilon V_2, \dots) \big| _{\varepsilon=0}.
\]
Here $H_{*}$ is a linear differential operator of order $k$ with coefficients in~$\mathcal O$. For example, $(U_2+U U_1)_{*}=D^2+L_U D+R_{U_1}$.

The definition of conserved density has to be modified. 
In the scalar case~\cite{Olv93} conserved densities are defined up to total $x$-derivatives, i.e.~two conserved densities are equivalent if~$\rho_1-\rho_2\in{\mathcal F}/\operatorname{Im}D$.
In the matrix case the conserved densities are traces of some matrix polynomials defined up to total derivatives. 

In the non-abelian case $\rho_{1} \sim \rho_{2}$ iff $\rho_{1} - \rho_{2}\in \mathcal{A}/({\rm Im}\,D+[\mathcal{A},\, \mathcal{A}])$.
The equivalence class of an element $\rho$ is called the {\it trace of} $\rho$ and is denoted by $\operatorname{tr}\rho$.

\begin{definition} The equivalence class of an element $\rho\in \mathcal{A}$ is called {conserved density} for equation~\eqref{nonabel} if $D_t(\rho)\sim 0$.
\end{definition}

Poisson brackets~\eqref{PDEbrop} are defined on the vector space $\mathcal{A}/({\rm Im}\,D+[\mathcal{A},\, \mathcal{A}])$.
A general theory of Poisson and double Poisson brackets on algebras of differential functions was developed in~\cite{SKV}.  
An algebra of (non-commutative) differential functions is defined as a unital
associative algebra $\mathcal{D}$ with a derivation $D$ and commuting derivations $\partial_i, \,\, i\in \Z_{+}$
such that the following two properties hold:
\begin{enumerate}[label=\arabic*),itemsep=1pt]
\item For each $f\in \mathcal{D}$ , $\partial_i(f)=0 $ for all but finitely many $i$;
\item $[\partial_i,\, D]=\partial_{i-1}$.
\end{enumerate}

\paragraph{Formal symmetry.}
 
At least for non-abelian equations of the form~\eqref{nonabel}, where
\begin{equation}\label{cononabel}
F=U_{n}+f(U,\, U_{1},\, \dots, \,U_{n-1})
\end{equation}
all definitions and results concerning formal symmetries
(as in Subsection~\ref{SectionIntegrabilityConditions}) can be easily generalized.

\begin{definition} A formal series
\[
\Lambda=D+l_{0}+l_{-1}D^{-1}+\cdots \, ,
\qquad l_k \in \mathcal{O}
\]
is a {\it formal symmetry} of order~1 for an equation~\eqref{cononabel}
if it satisfies the equation
\[
D_t (\Lambda)-[F_{*},\,\Lambda]=0\, .
\]
\end{definition}
For example, for the non-abelian Korteweg-de Vries equation
\eqref{matkdv} one can take~$\Lambda=\mathcal{R}^{1/2},$ where $\mathcal{R}$ is the following recursion
operator for~\eqref{matkdv} (see~\cite{OlvSok98}):
\[
\mathcal{R}=D^2+2(L_U+R_U)+(L_{U_{x}}+R_{U_{x}})\,D^{-1}+
(L_U-R_U)\,D^{-1}\,(L_U-R_U)\, D^{-1}.
\]
There are analogues to Theorems \ref{tlsym}, \ref{svsok} in the non-commutative case.

\section{Non-evolutionary systems}

Consider non-evolutionary equations of the form
\begin{equation}\label{eq:eqnev}
u_{tt}=F(u,u_1,\ldots,u_{n};u_{t},u_{1t},\ldots,u_{mt})
\end{equation}
where $u_{mt}=\partial^{m+1}u/\partial x^m\partial t$. In this section~$\mathcal F$ denotes the field of functions of variables~$u,u_1,u_{2}, \dots$ and~$u_t,u_{1t},u_{2t},\ldots$. We will say that equation~\eqref{eq:eqnev} is of order~$(n,m)$. The total $x$-derivative~$D$ and $t$-derivative~$D_t$ are
\[D=\sum_{i=0}^{\infty}u_{i+1}\frac{\partial}{\partial u_i}+\sum_{j=0}^{\infty}u_{j+1t}\frac{\partial}{\partial u_{jt}},\qquad D_t= \sum_{i=0}^\infty u_{it} \frac{\partial}{\partial u_i}
+\sum_{j=0}^\infty D^{j}(F) \frac{\partial}{\partial u_{jt}}  
\]  
and commute. It is easy to see that~$\operatorname{Ker}D=\C$ in~$\mathcal{F}$.

Evolutionary vector fields that commute with~$D$, can be written as
\[D_H= \sum_{i=0}^\infty D^i(H)\frac{\partial}{\partial u_i}
+\sum_{j=0}^\infty D^{j}(D_t(H)) \frac{\partial}{\partial u_{jt}}.
\]
For any function~$H(u,u_1,\ldots,u_t,u_{1t},\ldots)\in \mathcal{F}$ the differential operator $H_*$ is defined as 
\[H_*=\sum_{i=0}^\infty \frac{\partial H}{\partial u_i}D^i
+\sum_{j=0}^\infty\frac{\partial H}{\partial u_{jt}}D^{j}D_t.
\]
The linearization operator (see Section \ref{sec1.1}) for~\eqref{eq:eqnev} has the form\marginpar{\hspace{1sp}}
\[
\mathcal{L} = D_t^2 - F_* = D_t^2 - U - V\, D_t,
\]
where the differential operators $U$ and $V$ 
\begin{equation}\label{eq:symbs}
\begin{aligned}
U&= \uu_{n}D^{n}+\uu_{n-1}D^{n-1}+\cdots+\uu_{0},\\[2mm]
V&= \vv_{m}D^{m}+\vv_{m-1}D^{m-1}+\cdots+\vv_{0}
\end{aligned}\quad\text{with}\quad
\uu_{i}= \frac{\partial F}{\partial u_{i}},\quad \vv_{j}= \frac{\partial F}{\partial u_{jt}}
\end{equation}
are defined by the rhs of~\eqref{eq:eqnev}.

\begin{remark} We can rewrite~\eqref{eq:eqnev} in an evolutionary form as
\begin{equation}\label{eq:eqsys}
u_t=v,\qquad 
v_t=F(u,u_1,\ldots,u_r,v,v_x,\ldots,v_s). 
\end{equation}
The matrix linearization operator for~\eqref{eq:eqsys} is $D_t - \mathbf{F}$, where 
\[
\mathbf{F} = \mc{0&1\\U&V}.
\]
The approach in~\cite{MikSha85,MikSha86} is not applicable in the classification of integrable systems~\eqref{eq:eqsys}, since it requires the diagonalizability of the matrix differential operator~$\mathbf{F}$, and here it is not diagonalizable.  Assuming polynomiality of the equations, a powerful symbolic technique has been developed and applied to perform classifications of integrable systems~\eqref{eq:eqnev} in~\cite{MNW,NovWang}. 
\end{remark}

 It this paper we develop the approach proposed in~\cite{HSS}, which does not assume the polynomiality of the equations.

\subsection{Formal recursion operators}

\begin{definition}
The pseudo-differential operator $\mathcal{R}=X+Y\,D_{t}$ with components
\[
 X= \sum_{-\infty}^{p} x_{i} D^{i}, \qquad
 Y= \sum_{-\infty}^{q} y_{i} D^{i}
\]
is called
{\it formal recursion operator of order $(p,q)$} for
equation~\eqref{eq:eqnev} if it satisfies the relation 
\begin{equation} \label{defrec}
\mathcal{L} (X+Y\,D_{t})=(\bar X+\bar Y\,D_{t}) \mathcal{L}
\end{equation}
for some formal series $\bar X$, $\bar Y$.
\end{definition}
It follows from~\eqref{defrec} that $\bar Y=Y$ and $\bar X=X+2 Y_{t}+[Y,\, V]$.
If $X$ and $Y$ are differential operators
(or ratios of differential operators), condition~\eqref{defrec}
implies the fact that operator~$\mathcal{R}$ maps symmetries of
equation~\eqref{eq:eqnev} to symmetries. 

\begin{lemma}\label{lem4} Relation~\eqref{defrec} is equivalent to the identities
\be\label{*}
X_{tt}-V X_t+[X,U]+(2Y_t+[Y,V])U+YU_t=0,
\ee
\be\label{**}
Y_{tt}+2X_t+[Y,U]+[X,V]+([Y,V]+2Y_t)V+YV_t-VY_t=0.
\ee
\end{lemma}

Let $\mathcal{R}_{1}=X_{1}+Y_{1}\,D_{t}$ and $\mathcal{R}_{2}=X_{2}+Y_{2}\,D_{t}$
be two formal recursion operators. Then the product
$\mathcal{R}_{3}=\mathcal{R}_{1} \mathcal{R}_{2}$, in which $D_{t}^{2}$ is replaced by
$(U+V D_t)$ is also a formal recursion operator whose components are
given by
\[
 X_3=X_1X_2+Y_1Y_2 U+Y_1(X_{2})_t,\qquad Y_3=X_1Y_2+Y_1X_2+Y_1Y_2 V+Y_1(Y_{2})_t.
 \]
Thus the set of all formal recursion operators forms an associative algebra $A_{\text{frec}}$. In the evolution case this algebra is generated by one generator  of the form~\eqref{FSym}. For equation of the form~\eqref{eq:eqnev} the structure of $A_{\text{frec}}$ essentially depends on the numbers~$n$ and $m$.

\begin{definition}
The pseudo-differential operator $\mathcal{S}=P+Q\,D_{t}$ is called 
{\it formal symplectic operator} for
equation~\eqref{eq:eqnev} if it satisfies the relation\footnote{In the evolution case $\mathcal{L}=D_t-F_*$ and~\eqref{defsec} coincide with~\eqref{Req}.} 
\begin{equation} \label{defsec}
\mathcal{L}^+ (P+Q\,D_{t})+(\bar P+\bar Q \,D_{t}) \mathcal{L}=0
\end{equation}
for some formal series $\bar P, \bar Q$.
\end{definition}

The operator equations for the components of a formal symplectic operator~$\mathcal{S}$ have the following form
\begin{align} 
\label{eqcon1}
P_{tt}+V^{*} P_{t}+2 Q_{t} U+Q U_{t}&=U^{*} P-P U-(QV+V^{*}Q)\,U-V_{t}^{*}P,\\[2mm]
\label{eqcon2}
\begin{split}
Q_{tt}+2 P_{t}+2 Q_{t} V+V^{*} Q_{t}&=U^{*} Q-Q U-(QV+V^{*}Q)\,V
\\
&\qquad\qquad{}-(PV+V^{*}P)-(V_{t}^{*}Q+Q V_{t})
\end{split}
\end{align}
and
\[\bar{P}=-P-2Q_t-V^+Q-QV,\qquad \bar{Q}=-Q.
\]

\begin{definition}\label{def:fintnevol} We call equation~\eqref{eq:eqnev} {formally integrable} if it possesses a formal recursion operator $\mathcal R$ of some order $(p,q)$ with a complete set of arbitrary integration constants. 
\end{definition}

Since $\mathcal R$ depends on integration constants linearly, actually we have an infinite-dimensional vector space of formal recursion operators.

\subsection{Examples}

\paragraph{Example of order (3,1).}
\begin{example}\label{ex14}
The simplest example of integrable equation of the form~\eqref{eq:eqnev} is given by~\cite{luis3}:
\be\label{311}
u_{tt}=u_{xxx}+3u_x u_{xt}+(u_t-3 u_x^2)\,u_{xx}.
\ee
\end{example}
The formal recursion operator of order $(n+1,n)$\footnote{We do not assume that the leading coefficients are non-zeros and therefore we may postulate this order without loss of generality.} for this equation has the following structure on higher order terms
\begin{multline*}
(l_{n+1}+m_{n}nu_{x})\,D^{n+1}+\left[l_{n}+m_{n-1}(n{-}1)u_{x}+l_{n+1}(n{+}1)\left(\tfrac{1}{2}(n{-}4)u_{x}^{2}+u_{t}\right)\right.
\\
\left.+m_{n}\left(\tfrac{1}{6}(n{+}1)\left(n^{2}{-}7n{-}6\right)u_{x}^{3}+n(n{+}1)u_{t}u_{x}+\tfrac{1}{2}\left(n^{2}{-}n{+}2\right)u_{xx}\right)\right]\,D^{n}+\cdots
\\
{}+\left\{m_{n}D^{n}+\left[m_{n-1}+l_{n+1}(n{+}1)u_{x}+m_{n}(n{+}1)\left(\tfrac{n}{2}u_{x}^{2}+u_{t}\right)\right]\,D^{n-1}+\cdots\right\} D_{t}
\end{multline*}
where $m_i, l_i$ are the integration constants. Denote by 
$\mathcal{Y}_i$ the recursion operator corresponding to $m_i = 1$, $m_j=0$ when $j\ne i$, and $l_j=0$, and denote by~$\mathcal{X}_i$ the recursion operator corresponding to~$l_i=1$, $l_j=0$ if~$j\neq i$ and~$m_j=0$. The operator $\mathcal{Y}_{-1}$ can be written in an explicit form as
\[
\mathcal{Y}_{-1} = -u_{x} + D^{-1} (D_{t}+2 u_{xx}).
\]
We have that
\[
\mathcal{Y}_{-1}^{-1}=\mathcal{Y}_{-2},\quad\mathcal{Y}_{-1}^{2}=\mathcal{X}_{1},\quad\mathcal{X}_{1}^{-1}=\mathcal{X}_{-1},\quad\mathcal{X}_{1}^{i}=\mathcal{X}_{i}.
\]
Therefore any recursion operator $\mathcal{R}$ can be uniquely represented in the form
\[
\mathcal{R} = \sum_{-\infty}^{k} c_i \mathcal{Y}_{-1}^{i}, \qquad c_i\in \C.
\]
This equation does not admit any formal symplectic operator (see Lemma \ref{lem6}).

\paragraph{The potential Boussinesq equation.} 

\begin{example}
An example of integrable equation of order~(4,1) is the potential Boussinesq equation
\begin{equation}\label{eq:bouss}
u_{tt}=u_{xxxx}-3u_xu_{xx}.
\end{equation}
\end{example}
Similar to Example \ref{ex14},  we can find two different types of formal recursion operators:
\begin{multline}\label{eq:eqXi}
\mathcal{Y}_i=-\tfrac{3}{8} (i+2) u_{t}\, D^{-1+i}-\tfrac{3}{16} (i-2) (i+1) u_{xt}\, D^{-2+i}
\\
+\tfrac{1}{32} \left[9 (i-1) (i+2) u_{x} u_{t}-2 i \left(i^2-3 i+8\right) u_{xxt}\right]\, D^{-3+i} +\cdots
\\
{}+\left\{D^{i}-\tfrac34 i u_{x}\, D^{-2+i}-\tfrac{3}{8} (i-2) i u_{xx}\, D^{-3+i}
\right.
\\
\left.{}+ \left[\tfrac{9}{32}(i^2 -3 i +2) u_{x}^2-\tfrac{1}{16}(i{-}3) (i{-}2) (2 i{+}1) u_{xxx}\right]\, D^{-4+i}+\cdots
\right\}D_t
\end{multline}
and
\begin{multline}\label{eq:eqYi}
\mathcal{X}_i=D^{i}-\tfrac{3}{4} i u_{x}\, D^{-2+i}-\tfrac{3}{8} (i-2) i u_{xx}\, D^{-3+i}
\\
{}+\tfrac{1}{32} \left[9 (i-3) i u_{x}^2-2 (2i-7) (i-1) i u_{xxx}\right]\, D^{-4+i}+\cdots
\\
{}+\left\{-\tfrac{3}{8} i u_{t}\, D^{-5+i}-\tfrac{3}{16} (i-5) i u_{xt}\, D^{-6+i}
\right.
\\
\left.
{}+\left[\tfrac{9}{32} (i-5) i u_{x} u_{t}-\tfrac{1}{16} (i-5) (i-4) i u_{xxt}\right]\, D^{-7+i}+\cdots
\right\}\,D_t. 
\end{multline}
However, algebraic relations between them are completely different:
\[\mathcal{X}_i=\mathcal{X}_1^i,\qquad\mathcal{Y}_i=\mathcal{Y}_1\mathcal{X}_1^{i-1}.
\]
We also have that
\[\mathcal{Y}_1^2=\mathcal{X}_6,\quad\mathcal{Y}_1^{-1}=\mathcal{Y}_{-5}.
\]
so, notably, $\mathcal{Y}_{-2}^2=1$. We see that the algebra of all formal recursion operators is generated by $\mathcal{X}_{1}$ and $\mathcal{Y}_{1}.$ The generators commute with each other\footnote{A proof that these statements are true uses the homogeneity of~\eqref{eq:bouss} and~\eqref{eqcon1},~\eqref{eqcon2}.} and are related by the algebraic curve 
\[
\mathcal{X}_{1}^2 = \mathcal{Y}_{1}^6.
\]
The operator $\mathcal{Y}_{1}$ can be written in closed form as
\begin{multline}\label{eq:rcx1}
\mathcal{Y}_{1}=-\tfrac98u_t+\tfrac38D^{-1}\cdot u_{xt}+\left[D-\tfrac38u_xD^{-1}-\tfrac38D^{-1}\cdot u_x\right]D_t
\\
=-\tfrac98u_t+DD_t+\tfrac38D^{-1}\cdot\left( u_{xt}-u_xD_t\right)-\tfrac38u_xD^{-1}D_t.
\end{multline}
The independent classical symmetries of~\eqref{eq:bouss} are $1$, $s_0^a=u_x$ and~$s_0^b=u_t$. Applying the recursion operator $\mathcal{Y}_{1}$ to them, we obtain a chain of symmetries:
\begin{align*}
s^{\rm b}_1&=\mathcal{Y}_1(u_x)=u_{xxt}-\tfrac32u_xu_t,
\\[1.5mm]
s^{\rm a}_2&=\mathcal{Y}_1(u_t)=u_{xxxxx}-\tfrac{15}{4} u_{x} u_{xxx}-\tfrac{15}{16}u_{t}^2-\tfrac{45}{16} u_{xx}^2+\tfrac{15}{16}u_{x}^3.
\end{align*}
In general,~$\mathcal{Y}_1(s^{\textrm{a}}_{i})=s^{\textrm{b}}_{i+1}$ and~$\mathcal{Y}_1(s^{\textrm{b}}_{i})=s^{\textrm{a}}_{i+2}$, $i=0,1,\ldots$ where the notation indicates that there are two types of symmetries 
\begin{equation*}
\begin{aligned}
s^{\textrm{a}}_{i}&=c_i\,u_{2i+1}+\text{lower order terms},\\
s^{\textrm{b}}_{i}&=d_i\,u_{2it}+\text{lower order terms},
\end{aligned}
\qquad i=0,1,\ldots
\end{equation*}
Notice that symmetries of type~$s^{\textrm{a}}_{1+3i}$ and~$s^{\textrm{b}}_{2+3i}$ with~$i=0,1,\ldots$ are not produced by this scheme. 
%The algebra of formal recursion operators can be in principle generated by two elements, $\mathcal{Y}_1$ and $\mathcal{X}_1$. But~$\mathcal{X}_1$ is a formal operator that cannot be written, apparently, in closed form, and we have only a recursive procedure to build it up. If this operator can be written in local form and commutes with~$\mathcal{Y}_1$, then the whole algebra of formal recursion operators can be generated through the relations
% Particularly, the first coefficients of~$\mathcal{Y}_{1}^{-1}$ are
%\begin{multline*}
%\mathcal{Y}_{1}^{-1}=
%\tfrac{9}{8}u_{t}\, D^{-6}-\tfrac{21}{4}u_{xt}\, D^{-7}+\tfrac{3}{16} \left(27 u_{x} u_{t}+80 u_{xxt}\right)\, D^{-8}
%\\
%{}-\tfrac{27}{16} \left(12 u_{t} u_{xx}+19 u_{x} u_{xt}+20 u_{xxxt}\right)\, D^{-9}+\cdots
%\\
%{}+\left[D^{-5}+\tfrac{15}{4}u_{x}\, D^{-7}-\tfrac{105}{8}u_{xx}\, D^{-8}+\tfrac{63}{16} \left(3 u_{x}^2+8 u_{xxx}\right)\, D^{-9}+\cdots\right]D_{t} 
%$\end{multline*}
 
\begin{remarknn} The potential Boussinesq  equation~\eqref{eq:bouss} is Lagrangian (see Subsection~\ref{Lagr}) and therefore  its linearization operator is self-adjoint: $\mathcal{L}^+=\mathcal{L}.$ For such equations recursion and symplectic operators coincide. In particular, the recursion operator~\eqref{eq:rcx1} is also a symplectic operator for equation~\eqref{eq:bouss}.
\end{remarknn}

%\paragraph{Summary.} 

%For any equation~\eqref{eq:eqnev} with~$2m<n$, the terms with highest order on~$D$ in~\eqref{*}--\eqref{**} are in~$[X,U]$ and~$[Y,U]$  and therefore the coefficients of a generic formal recursion operator can be found recursively from~\eqref{*} and~\eqref{**} by solving equations of the form $D(X) = f, \,\, f\in \mathcal{F}$.  Since $\operatorname{Ker}D=\C$, some integration constants appear of each step (cf.~Remark~\ref{rem24}).
 
%the series~$\mathcal{Y}_i$ and~$\mathcal{X}_i$ can be found recursively from relations~\eqref{*} and~\eqref{**}. 

\subsection{Integrability conditions}

For any equation~\eqref{eq:eqnev} with~$2m<n$, the terms with highest order on~$D$ in~\eqref{*}--\eqref{**} are in~$[X,U]$ and~$[Y,U]$ and produce equations where, using integrating factors of the form~$\mathfrak{u}_4^\alpha$, the unknowns~$x_{i}$,~$y_{j}$ can be isolated in two equations of the form
\begin{equation}\label{eq:eqsml}
D\left(\mathfrak{u}_{4}^{-i/4}x_{i}\right)=A_{i},\qquad
D\left(\mathfrak{u}_{4}^{-i/4}y_{i}\right)=B_{i}
\end{equation}
The right hand sides~$A_{i},$~$B_{i}$ depend only on variables~$x_j$ and~$y_j$ of lower indexes. Therefore the coefficients of the series~$X$ and~$Y$ can be found recursively from relations~\eqref{*} and~\eqref{**} but there appear an infinite number of integrability obstructions because equations~\eqref{eq:eqsml} have no solutions for arbitrary $A_i$ and $B_i$\footnote{these functions must be total derivatives.}.  
Since~$\operatorname{Ker}D=\C$, some integration constants arise from~\eqref{eq:eqsml} of each step (cf.~Remark~\ref{rem24}).

Proposition~\ref{Lambdaspace} implies that in the evolution case the obstructions to formal integrability do not depend on the choice of the integration constants and on the order of formal recursion operator. 

\begin{conjecture}\label{conj1} This is also true for equations~\eqref{eq:eqnev} with $2m<n$.  
\end{conjecture}

To find the coefficients of a formal symplectic operator we have to use the relations~\eqref{eqcon1},~\eqref{eqcon2}. An analysis 
of the terms with highest order on~$D$ leads (cf. Subsection \ref{ssFor}) to the following 
\begin{lemma}\label{lem6}  If~$2m<n$ and $n$ is odd, then no non-trivial formal symplectic operator exists.
\end{lemma}

Here we consider two classes of equations~\eqref{eq:eqnev}, of order~(3,1) and of order~(4,1).  The computations justify the conjecture in these two cases.

\subsubsection{Equations of order $(3,1)$.} 
In the paper~\cite{HSS} equations of the form 
\begin{equation}\label{eq:eqs31g}
u_{tt}=f(u_{3},u_{1t},u_{2},u_{t},u_1,u),\qquad\frac{\partial f}{\partial u_{3}}\neq0
\end{equation}
were considered. According to Lemma \ref{lem6}, such equations have no formal symplectic operator.

The first integrability conditions for the existence of a formal recursion operator have the form of conservation laws $D_t(\rho_i)=D(\sigma_i)$, $i=0,1,\dots$ (see~\cite{HSS,CH}) and the conserved densities can be written as
\begin{gather*}
\rho_{0}=\frac{1}{\sqrt[3]{\mathfrak{u}_{3}}},\qquad
\rho_{1}=\frac{3 \mathfrak{u}_2}{\mathfrak{u}_3}+\frac{2\sigma_1}{\sqrt[3]{\mathfrak{u}_3}}+\frac{\sigma_0\mathfrak{v}_1}{\mathfrak{u}_3^{2/3}}+\frac{\sigma_0^2}{\sqrt[3]{\mathfrak{u}_3}},\qquad
\rho_{2}=\frac{\mathfrak{v}_1}{\mathfrak{u}_3^{2/3}}-\frac{2\sigma_0}{\sqrt[3]{\mathfrak{u}_3}},
\\[2mm]
\begin{split}
\hspace{-7mm}\rho_{3}=\frac{\mathfrak{v}_1^3}{\mathfrak{u}_3^{4/3}}-\frac{27 \mathfrak{v}_0}{\sqrt[3]{\mathfrak{u}_3}}+\frac{9 \mathfrak{u}_2 \mathfrak{v}_1}{\mathfrak{u}_3^{4/3}}-\frac{9 \mathfrak{v}_1 D\mathfrak{u}_3}{\mathfrak{u}_3^{4/3}}-\frac{6 \sigma_2}{\sqrt[3]{\mathfrak{u}_3}}+\frac{3 \sigma_1 \mathfrak{v}_1}{\mathfrak{u}_3^{2/3}}+\frac{6 \sigma_1 \sigma_0}{\sqrt[3]{\mathfrak{u}_3}}
+\frac{3\sigma_0^2 \mathfrak{v}_1}{\mathfrak{u}_3^{2/3}}
+\frac{4 \sigma_0^3}{\sqrt[3]{\mathfrak{u}_3}},
\end{split}
\\[2mm]
\begin{split}
\hspace{-7mm}\rho_{4}=-\frac{81 \mathfrak{u}_1}{\mathfrak{u}_3^{2/3}}+\frac{27 \mathfrak{u}_2^2}{\mathfrak{u}_3^{5/3}}+\frac{\mathfrak{v}_1^4}{\mathfrak{u}_3^{5/3}}+\frac{9 \mathfrak{u}_2 \mathfrak{v}_1^2}{\mathfrak{u}_3^{5/3}}-\frac{27 \mathfrak{v}_0 \mathfrak{v}_1}{\mathfrak{u}_3^{2/3}}-\frac{27 \mathfrak{u}_2 D\mathfrak{u}_3}{\mathfrak{u}_3^{5/3}}-\frac{9 \mathfrak{v}_1^2 D\mathfrak{u}_3}{\mathfrak{u}_3^{5/3}}+\frac{9 \left(D\mathfrak{u}_3\right)^2}{\mathfrak{u}_3^{5/3}}
\\
+\frac{2 \sigma_3}{\sqrt[3]{\mathfrak{u}_3}}+\frac{3 \sigma_2 \mathfrak{v}_1}{\mathfrak{u}_3^{2/3}}+\frac{6 \sigma_2 \sigma_0}{\sqrt[3]{\mathfrak{u}_3}}-\frac{3 \sigma_1^2}{\sqrt[3]{\mathfrak{u}_3}}-\frac{6 \sigma_1 \sigma_0 \mathfrak{v}_1}{\mathfrak{u}_3^{2/3}}-\frac{12 \sigma_1 \sigma_0^2}{\sqrt[3]{\mathfrak{u}_3}}-\frac{7 \sigma_0^4}{\sqrt[3]{\mathfrak{u}_3}}-\frac{5 \sigma_0^3 \mathfrak{v}_1}{\mathfrak{u}_3^{2/3}}
\\
-\frac{\sigma_0 \mathfrak{v}_1^3}{\mathfrak{u}_3^{4/3}}+\frac{27 \sigma_0 \mathfrak{v}_0}{\sqrt[3]{\mathfrak{u}_3}}-\frac{9 \sigma_0 \mathfrak{u}_2 \mathfrak{v}_1}{\mathfrak{u}_3^{4/3}}
+\frac{9 \sigma_0 \mathfrak{v}_1 D\mathfrak{u}_3}{\mathfrak{u}_3^{4/3}}-27 D_{t}\left(\frac{\sigma_0}{\sqrt[3]{\mathfrak{u}_3}}\right)-\frac{54 \mathfrak{v}_1 D\sigma_0}{\sqrt[3]{\mathfrak{u}_3}}.
\end{split}
\end{gather*}

\subsubsection{Equations of order $(4,1)$.} 
Similar conditions for integrable equations of the form 
\begin{equation}\label{eq:eqs41g}
u_{tt}=f(u_{4}, u_{3},u_{1t},u_{2},u_{t},u_1,u),\qquad\frac{\partial f}{\partial u_{4}}\neq0
\end{equation}
are given~\cite{CH} by the conserved densities:
\begin{gather*}
\rho_{0}=\frac{1}{\sqrt[4]{\mathfrak{u}_{4}}},\qquad\rho_{1}=\frac{\mathfrak{u}_{3}}{\mathfrak{u}_{4}},\qquad\rho_{2}=\frac{\mathfrak{v}_{1}}{\sqrt{\mathfrak{u}_{4}}}-2\frac{\sigma_{0}}{\sqrt[4]{\mathfrak{u}_{4}}},
\\[2mm]
\rho_{3}=
-\frac{4\mathfrak{v}_{0}}{\sqrt[4]{\mathfrak{u}_{4}}}
+\frac{\mathfrak{u}_{3}\mathfrak{v}_{1}}{\mathfrak{u}_{4}^{5/4}}-\frac{3\mathfrak{v}_{1}D\mathfrak{u}_{4}}{2\mathfrak{u}_{4}^{5/4}}-\frac{2\sigma_{1}}{\sqrt[4]{\mathfrak{u}_{4}}},
\\[2mm]
\rho_{4}=-\frac{32\mathfrak{u}_{2}}{\mathfrak{u}_{4}^{3/4}}-\frac{4\mathfrak{v}_{1}^{2}}{\mathfrak{u}_{4}^{3/4}}+\frac{12\mathfrak{u}_{3}^{2}}{\mathfrak{u}_{4}^{7/4}}+\frac{5\left(D\mathfrak{u}_{4}\right)^{2}}{\mathfrak{u}_{4}^{7/4}}-\frac{12\mathfrak{u}_{3}D\mathfrak{u}_{4}}{\mathfrak{u}_{4}^{7/4}}-\frac{16\sigma_{2}}{\sqrt[4]{\mathfrak{u}_{4}}}-\frac{16\sigma_{0}^{2}}{\sqrt[4]{\mathfrak{u}_{4}}},
\\[2mm]
\rho_{5}=
\frac{16 \mathfrak{u}_1}{\sqrt{\mathfrak{u}_4}}
+\frac{3 \mathfrak{v}_1^2 D\mathfrak{u}_4}{\mathfrak{u}_4^{3/2}}
-\frac{3 \mathfrak{u}_3^2 D\mathfrak{u}_4}{\mathfrak{u}_4^{5/2}}
-\frac{2 \mathfrak{u}_3 \left(D\mathfrak{u}_4\right)^2}{\mathfrak{u}_4^{5/2}}
+\frac{4 \mathfrak{u}_3 D^2\mathfrak{u}_4}{\mathfrak{u}_4^{3/2}}
\\
{}+\frac{8 \mathfrak{u}_2 D\mathfrak{u}_4}{\mathfrak{u}_4^{3/2}}
-\frac{8 \mathfrak{u}_2 \mathfrak{u}_3}{\mathfrak{u}_4^{3/2}}
+\frac{8 \mathfrak{v}_0 \mathfrak{v}_1}{\sqrt{\mathfrak{u}_4}}
+\frac{2 \mathfrak{u}_3^3}{\mathfrak{u}_4^{5/2}}
-\frac{2 \mathfrak{u}_3 \mathfrak{v}_1^2}{\mathfrak{u}_4^{3/2}}
-\frac{2 \sigma_3}{\sqrt[4]{\mathfrak{u}_4}}
-\frac{4 \sigma_0 \sigma_1}{\sqrt[4]{\mathfrak{u}_4}}
\\
{}+8 D_{t}\left(\frac{\sigma_0}{\sqrt[4]{\mathfrak{u}_4}}\right)
+\frac{16 \mathfrak{v}_1 \left(D\sigma_0\right)}{\sqrt[4]{\mathfrak{u}_4}}
-\frac{3 \sigma_0 \mathfrak{v}_1 \left(D\mathfrak{u}_4\right)}{\mathfrak{u}_4^{5/4}}
+\frac{2 \sigma_0 \mathfrak{u}_3 \mathfrak{v}_1}{\mathfrak{u}_4^{5/4}}-\frac{8 \sigma_0 \mathfrak{v}_0}{\sqrt[4]{\mathfrak{u}_4}}.
\end{gather*}
\begin{remarknn}
Supposing $X$ of order $n+1$ and $Y$ of order $n$ in~\eqref{*},~\eqref{**},  the density $\rho_0$ comes
from the coefficient of $D^{n+4}$ in~\eqref{*}, $\rho_1$ from the coefficient of $D^{n+3}$ in~\eqref{*}, $\rho_2$ from
$D^{n+1}$ in~\eqref{**}, $\rho_3$ from $D^n$ in~\eqref{**}, $\rho_4$ from $D^{n+2}$ in~\eqref{*}, $\rho_5$ from $D^{n+1}$ in~\eqref{*}, $\rho_6$ from $D^{n-1}$ in~\eqref{**}, $\rho_7$ from $D^{n-2}$ in~\eqref{**}, etc.
\end{remarknn}

It would be interesting to find a recurrent formula for these conserved densities similar to~\eqref{rekkur_sc}.

The existence of a formal symplectic operator imposes additional conditions (cf.~Theorem \ref{contriv}). The first calculated conditions are of a similar nature than in the evolution case: the densities~$\rho_i$ must be conserved and, additionally, some of them must be trivial, i.e.~total derivatives. Concretely, at the time of writing this report we know that~$\rho_1=D\omega_1$, $\rho_3=D\omega_3$, $\rho_5=D\omega_5$, and~$\rho_7=D\omega_7$,
where $\omega_i\in\mathcal{F}$ are local functions. 

The general formulas for the first coefficients of formal recursion  operators for a general equation~\eqref{eq:eqs41g}, written  in terms of coefficients of the linearization operator (see~\eqref{eq:symbs}), are given by 
 
\begin{multline}
\mathcal{X}_i=\mathfrak{u}_4^{\frac{i}{4}} D^{i}+\left[\tfrac{1}{4} i \mathfrak{u}_3 \mathfrak{u}_4^{\frac{i}{4}-1}+\tfrac{1}{8} (i-4) i \left(D\mathfrak{u}_4\right) \mathfrak{u}_4^{\frac{i}{4}-1}\right] D^{-1+i}+\cdots
\\
{}+\left\{\left[\tfrac{1}{4} i \mathfrak{u}_4^{\frac{i}{4}-1} \mathfrak{v}_1-\tfrac{1}{2} i \sigma_0 \mathfrak{u}_4^{\frac{i}{4}-\frac{3}{4}}\right]\, D^{-3+i}+\cdots\right\}D_t
\end{multline}
\begin{multline}
\mathcal{Y}_i=\left[\tfrac{1}{4} i \mathfrak{u}_4^{\frac{i}{4}} \mathfrak{v}_1-\tfrac{1}{2} (i+2) \sigma_0 \mathfrak{u}_4^{\frac{i}{4}+\frac{1}{4}}\right] D^{1+i}+\cdots
\\
{}+\left\{
\mathfrak{u}_4^{\frac{i}4} D^{i}+\left[\tfrac{1}{4} i \mathfrak{u}_3 \mathfrak{u}_4^{\frac{i}{4}-1}+\tfrac{1}{8} (i-4) i \left(D\mathfrak{u}_4\right) \mathfrak{u}_4^{\frac{i}{4}-1}\right] D^{-1+i}+\cdots\right\}D_t.
\end{multline}
The next coefficients depend on the fluxes $\sigma_i$, which correspond to the canonical conserved densities $\rho_i$ shown above. The integration constants are hidden in the fluxes.

The formal symplectic operators have the following structure: 
\begin{multline}
\mathcal{P}_i=
\kappa \mathfrak{u}_4^{\frac{i}{4}-1}D^{i}+\kappa\left[\tfrac{1}{4} i \mathfrak{u}_3 \mathfrak{u}_4^{\frac{i}{4}-2}+\tfrac{1}{8} (i-4) i (D\mathfrak{u}_4)
\mathfrak{u}_4^{\frac{i}{4}-2}\right]D^{-1+i}
+\cdots
\\
{}+\left\{\kappa\left[\tfrac{1}{4} (i{-}1)  \mathfrak{v}_1 \mathfrak{u}_4^{\frac{i}{4}-2}-\tfrac{1}{8} \omega_3 \mathfrak{u}_4^{\frac{i}{4}-\frac{7}{4}}-\tfrac{1}{2} (i{-}4) \sigma_0 \mathfrak{u}_4^{\frac{i}{4}-\frac{7}{4}}\right] D^{-3+i}
+\cdots\right\}D_t
\end{multline}
\begin{multline}
\mathcal{Q}_i= \kappa \left[\tfrac{1}{4} (i-1)\mathfrak{v}_1 \mathfrak{u}_4^{\frac{i}{4}-1}-\tfrac{1}{8} \omega_3 \mathfrak{u}_4^{\frac{i}{4}-\frac34}-\tfrac{1}{2} (i-2) \sigma_0 \mathfrak{u}_4^{\frac{i}{4}-\frac34}\right] D^{1+i}+\cdots
\\
{}+\left\{
\kappa \mathfrak{u}_4^{\frac{i}{4}-1}\, D^{i}+\kappa\left[\tfrac{1}{4} i \mathfrak{u}_3 \mathfrak{u}_4^{\frac{i}{4}-2}+\tfrac{1}{8} (i{-}4) i\left(D\mathfrak{u}_4\right) \mathfrak{u}_4^{\frac{i}{4}-2}\right] D^{-1+i}
+\cdots\right\}D_t
\end{multline}
where the function $\kappa$ is defined by the formula $D\kappa=2\mathfrak{u}_3\kappa/\mathfrak{u}_4$. Since the density $\rho_1$ is trival, $\kappa\in \mathcal{F}.$

\subsection{Lists of integrable equations}

\subsubsection{Equations of order $(3,1)$.} 
In~\cite{HSS} we find that the only integrable equations of type 
\begin{equation}\label{eq:eqs31}
u_{tt}=u_{3}+f(u_{1t},u_{2},u_t,u_1,u).
\end{equation}
up to certain point transformations are
\begin{align}
\begin{split}\label{eqq1}
u_{tt}&=u_{3}+(3u_1+k)u_{1t}+(u_t-u_1^2-2ku_1+6\wp)u_2-2\wp'u_t+6\wp'u_1^2
\\
&\qquad{}+(\wp''+k\wp')u_1,
\end{split}
\\[2.5mm]
\begin{split}\label{eqq2}
 u_{tt}&=u_3+\left[3\frac{u_t}{u_1}+\tfrac32X(u)\right]u_{1t}-\frac{u_2^2}{u_1}-\left[3\frac{u_t^2}{u_1^2}+\tfrac32X(u)\frac{u_t}{u_1}\right]u_2
\\&\qquad{}+c_2\left[u_1u_t+\tfrac32X(u)u_1^2\right],
\end{split}
\end{align}
where $\wp=\wp(u)$ is a solution of~$(\wp')^2=8\wp^3+k^2\wp^2+k_1\wp+k_0$, $X(u)=c_2u+c_1$ and~$k_1,k_2,c_1,c_2$ are arbitrary constants. Both are linearizable.

\subsubsection{Lagrangian equations of order $(4,1)$.}\label{Lagr} 
In~\cite{CH}, a classification of integrable Lagrangian systems with Lagrangians of the form
\[
L=\frac12 L_2(u_{2}, u_1, u)\,u_t^2 + L_1(u_{2}, u_1, u)\, u_{t} + L_0(u_{2}, u_1, u)
\]
was performed. The corresponding Euler-Lagrange equations are of the form~\eqref{eq:eqs41g}, and a total of eight systems were found, with Lagrangians:
\begin{gather}
\label{eq:lagrD1f}
L_1=\frac{u_t^2}{2}
   +\epsilon\, u_1u_t
   +\frac{u_2^2}{2} + \delta_2\frac{u_1^2}{2} + \delta_1\frac{u^2}{2}+\delta_0 u,
\\
\label{eq:lagrD2f}
L_2=\frac{u_t^2}{2}
   -u_1^2u_t
   +\frac{u_2^2}{2}+\frac{u_1^4}{2},
\\
\label{eq:lagrD8f}
L_3=\frac{u_t^2}{2}
+\frac{u_2^2}2
+\frac{u_1^3}2,
\\
\label{eq:lagrD3f}
L_4=\frac{u_t^2}{2}+a(u)\,u_1u_t+\frac{u_2^2}{2u_1^4}+a'(u)u_1\log{u_1}+\frac{a^2(u)}{2}u_1^2+d(u),\\
\label{eq:lagrD5f}
L_5=\frac{u_t^2}{2}+\left(\frac{\gamma}{u_1}+\epsilon\, u_1\right)u_t+\frac{u_2^2}{2u_1^4}+\frac{\epsilon^2}{2}u_1^2+\frac{\gamma^2}{2u_1^2}+\frac{\delta}{u_1},\quad|\gamma|+|\delta|\neq0,
\\
\label{eq:lagrD6f}
L_6=\frac{u_1}2u_t^2+(\epsilon u_1+\beta)u_1u_t+\frac{u_2^2}{2u_1^3}+\frac{\epsilon^2}2u_1^3+\epsilon\beta\, u_1^2+\frac{\delta}{u_1},
\\
\label{eq:lagrD7f}
L_7=\frac{u_t^2}{2u_1}+\frac{b(u)}{u_1}\,u_t+\frac{u_2^2}{2a(u)^4u_1^5}
+\frac{d_2(u)}{u_1},
\\
\label{eq:lagrD10f}
L_8=\frac{u_t^2}{2(u_1^2-1)}{ + }\frac{u_2^2}{2(u_1^2-1)} {+} d(u)u_1^2{-}\frac{d(u)}{3},\quad
d'''(u)-8d(u)d'(u)=0.
\end{gather}
The Greek letters are arbitrary constants, and the Lagrangians are nonequivalent through contact transformations and total derivatives. The integrability of~$L_4$ was proved only if~$a'(q)=0$, while a non-constant~$a(q)$ leads probably to a non-integrable equation. All the other Lagrangians were proved integrable by providing explicit quasi-local recursion operators~$\mathcal R$, all of them being of type~$\mathcal{Y}_i$ or~$\mathcal{X}_i$.

\subsection{Quasi-local recursion operators}

The linear case~\eqref{eq:lagrD1f} admits two recursion operators~$\mathcal{L}_1=D$ and~$\mathcal{X_0}=D_t$ that create a double chain of symmetries starting from~$u$.

The search of explicit recursion operators is again facilitated by the quasi-local ansatz~\eqref{anz} adapted to the non-evolutionary case:
\begin{equation}\label{neqla}
\mathcal{R}=\mathcal{D}+\sum_ks_kD^{-1}\cdot\mathcal{C}_k,
\end{equation}
where~$\mathcal{D}$ is a differential operator, the coefficients~$s_k$ are symmetries and~$\mathcal{C}_k$ the variational derivatives of conserved densities. The only difference with the evolution case is that the variational derivative not a function but a differential operator of the form $p D_t+q,\,\,p,q\in\mathcal{F}.$ Namely, if $\rho_* =P + Q D_t$ then 
\[
\frac{\delta \rho}{\delta u}= P^+(1) + Q^+(1) D_t. 
\]
\begin{example} Equation~\eqref{eqq1} has essentially only one non-trivial
conserved density given by
\begin{equation} \label{dens}
\rho=u_{t}-u_{x}^{2}+2 \wp(u).
\end{equation}
The explicit recursion operator for this equation can be written in the form
\[
\mathcal{R}=D+(u_{t}-2 u_{x}^{2}-k u_{x}+2 \wp)+u_{x} D^{-1}(D_{t}+
2 u_{xx}+2 \wp').
\]
In the non-local term the factor $u_x$ is a symmetry and the operator $D_{t}+
2 u_{xx}+2 \wp'$ is the variational derivative of the conserved density. Thus this recursion operator fits into the quasi-local ansatz~\eqref{neqla} . Symmetries of equation~\eqref{eqq1} can be constructed by applying this recursion operator to the seed symmetries $u_x$ and $u_t$.

Another quasi-local recursion operator for~\eqref{eqq1} has the form
\[
\bar{\mathcal{R}}=D_{t}+(u_{xx}-u_{x}^{3}-k u_{x}^{2}+6 \wp  u_{x}+k \wp +
\wp ')+u_{t} D^{-1}(D_{t}+2 u_{xx}+2 \wp').
\]
One can verify that
\[
\bar{\mathcal{R}}^{2}=\mathcal{R}^{3}+c_{2} \mathcal{R}^{2}+c_{1} \mathcal{R}+c_{0} 
\]
for some constants $c_i$. Recall that the situation is similar for the Krichever-Novikov equation (see Subsection \ref{recHam}).
\end{example}
The recursion operator~\eqref{eq:rcx1} for the potential Boussinesq equation~\eqref{eq:bouss} has the form~\eqref{neqla}, where the two non-local terms are generated by the symmetries $s_1= 1, s_2= u_x$ and by the conserved densities $\rho_1 = u_x u_t$ and $\rho_2 = u_t$.

Another example is system~\eqref{eq:lagrD7f} that admits for all the arbitrary functions involved, notably, a local differential recursion operator 
\[\mathcal{Y}_0=-D\cdot\frac{u_t}{u_{x}}+D_{t}.
\]
This operator generates a chain of symmetries starting from~$u_t$. Usually, the existence of a local recursion operator shows that the equation is linearizable.

%Here will follow your acknowledgements, if any ...

\strut\hfill

\noindent

\section*{Acknowledgments}
V. Sokolov acknowledges support from state assignment 
No 0033-2019-0006 and of the State Programme of the Ministry of Education and Science of the Russian Federation, project No 1.12873.2018/12.1. R.~Hern\'andez~Heredero acknowledges partial support from ``Ministerio de Econom\'{i}a y Competitividad'' (MINECO, Spain) under grant MTM2016-79639-P (AEI/FEDER, EU).

%    {\bf A Remark on the references:}
%    The references should be ordered alphabetically according to the
%authors in the list and the authors' initials should be after their surname without a
%comma, e.g. Ovsienko V and Roger C. Please see below.
%


\begin{thebibliography}{99}
%  
%
%\bibitem{Conte-1999}
%Conte R (Ed), {\it The Painlev\'e Property One Century Later}, Springer, New York, 1999. 
%
%\bibitem{Fu}
%Fuchs D B, {\it Cohomology of infinite-dimensional Lie
%algebras}, Plenum Publ. New York, 1986.
%
%
%\bibitem{OR2}
%Ovsienko V and Roger C,
%Deforming the Lie algebra of vector fields on~$S^1$ inside
%the Poisson algebra on $\dot T^*S^1$, {\it Comm. Math. Phys.} {\bf
%198}, 97--110, 1998.
%
%
%\bibitem{OR1}
%Ovsienko V and Roger C, Deforming the Lie algebra of vector
%fields on~$S^1$ inside the Lie algebra of pseudodifferential
%operators on $S^1$, {\it AMS Transl. Ser.~2}, (Adv. Math. Sci.)
%{\bf 194}, 211--227, 1999.
%
%
\bibitem{AbGal1} Abellanas L and Galindo A, Conserved densities for nonlinear evolution equations I. Even order Case, {\it J. Math. Phys.}, {\bf 20}(6), 1239--1243, 1979.

\bibitem{AbeGal83} Abellanas L and Galindo A,
Evolution equations with high order conservation laws. {\it J.
Math. Phys.}, {\bf 24}(3), 504--509, 1983.

\bibitem{adler} Adler M, On a trace functional for formal pseudo differential operators and the symplectic structure of the Korteweg de Vries type equations, {\it Invent. Math.}, {\bf 50}(3), 219-248, 1979.

\bibitem{AMS} Adler V~E, Marikhin V~G and  Shabat  A~B, Lagrangian chains and canonical Backlund transformations, {\it Theoret. and Math. Phys.}, {\bf 129}(2), 1448--1465, 2001.

\bibitem{ASY} Adler  V~E, Shabat  A~B and Yamilov R~I, Symmetry approach to the integrability problem, {\it  Theoret. and Math. Phys.}, {\bf 125}(3), 355-424, 2000.

\bibitem{CH} Caparr\'os Quintero A and Hern\'andez Heredero R, Formal recursion operators of integrable nonevolutionary equations and Lagrangian Systems. {\it J. Phys. A: Math. Theor.}, {\bf 51}, 385201, 2018.

\bibitem{CheLeeLiu79} Chen H~H, Lee Y~C and Liu C~S,
Integrability of nonlinear Hamiltonian systems by inverse
scattering method, {\it Phys. Scr.}, {\bf 20}(3-4), 490--492, 1979.

\bibitem{SKV} De Sole A, Kac V~G, and Valeri D, Double Poisson vertex algebras and non-commutative Hamiltonian equations,
{\it Adv. Math.}, {\bf 281}, 1025--1099, 2015.

\bibitem{demsok} Demskoy D~K and  Sokolov V~V,  On recursion
operators for elliptic models,  {\it Nonlinearity}, {\bf 21},
1253--1264, 2008.

\bibitem{Dorf} Dorfman I~Ya,   {\it Dirac Structures and Integrability of Nonlinear Evolution Equations}, John Wiley\&Sons, Chichester, 1993.

\bibitem{DSS}  Drinfel'd V~G, Svinolupov S~I and  Sokolov V~V, Classification of fifth order evolution equations
possessing infinite series of conservation laws, {\it Dokl. AN USSR,}, {\bf A10}, 7--10, 1985 (in Russian).

\bibitem{fokas}  Fokas A~S, A symmetry approach to exactly solvable evolution equations, {\it J. Math. Phys.}. {\bf 21}(6), 1318--1325, 1980.

\bibitem{Fokas} Fokas  A~S,  Symmetries and integrability, {\it Studies in Applied Mathematics}, {\b 77}(3),  253--299, 1987. 

\bibitem{Fuch} Fuchssteiner B., {\it Application of Hereditary Symmetries
to Nonlinear Evolution Equations}, {\em Nonlinear Anal.
Theory Meth. Appl.}, {\bf 3}, 849--862,  1979. 

\bibitem{GMSh} Gel'fand I~M, Manin Yu~I, and Shubin M~A, Poisson brackets and kernel of variational derivative in formal variational calculus, {\it Funct. Anal. Appl.},  {\bf 10}(4), 30--34, 1976.

\bibitem{hss1}  Hern\'andez Heredero R, Sokolov V~V, and  Svinolupov S~I, Toward the classification of third order integrable evolution equations, {\it J. Phys. A.} {\bf 13}, 4557--4568, 1994.

\bibitem{HSS}  Hern\'andez Heredero R, Sokolov V~V, and  Shabat A~B, A new class of linearizable equations. {\it J. Phys. A: Math. Gen.}, {\bf 36}, L605--L614, 2003.

\bibitem{her} Hern\'andez Heredero R, Classification of fully nonlinear integrable evolution equations of third order, {\it J. Nonlin. Math. Phys.},  {\bf 12}(4), 567--585, 2005.

\bibitem{ibshab} Ibragimov N~Kh and  Shabat A~B, Infinite Lie-B\"aklund algebras, {\it Funct. anal. appl.}, {\bf 14}(4), 313--315, 1980.

\bibitem{ibrag} Ibragimov N~H, {\it Transformation Groups Applied to Mathematical Physics Mathematics},  Dordrecht: D. Reidel, 1985.


\bibitem{kapl} Kaplansky I, {\it An introduction to differential algebra}, Paris, Hermann, 1957.

\bibitem{Kap82} Kaptsov, O~V, 
Classification of evolution equations with respect to conservation
laws, {\it Func. analiz i pril.}, {\bf 16}(1), 73--73, 1982.

\bibitem{Kau80} Kaup D~J,  On the inverse scattering problem for the cubic eigenvalue problem of the class~$\varphi_{xxx}+6 Q \phi_x+6 R \phi=\lambda \phi$,
{\it Stud. Appl. Math.}, {\bf 62}, 189--216, 1980.

\bibitem{KLV} Krasilchchik I~S,  Lychagin V~V  and Vinogradov A~M, {\it Geometry of jet spaces and nonlinear partial differential equations},  Series	(Adv. Stud. Contemp. Math., {\bf 1}, Gordon and Breach, New York, 1996.

\bibitem{kuper} Kupershmidt B~A, {\it KP or mKP}, Providence, RI : American Mathematical Society, 600 pp., 2000.

\bibitem{luis3} Mart\'\i nez Alonso L and Shabat A B, Towards a theory of differential constraints of a hydrodynamic hierarchy, {\it J. Non. Math. Phys.}, {\bf 10}, 229--242, 2003.

\bibitem{malnov} Maltsev  A~Ya and  Novikov  S~P, On the local systems Hamiltonian in the weakly non-local Poisson brackets, 
{\it Phys. D}, {\bf 156}(1-2), 53--80, 2001.

\bibitem{man} Manakov S~V., Note on the integration of Euler's
equations of the dynamics of an n-dimensional rigid body,  {\it Funct. Anal. Appl.}, {\bf 10}(4),  93--94, 1976.

\bibitem{march} Marchenko V~A, {\it Nonlinear equations and operator algebras}, Naukova dumka, Kiev, 152 pp., 1986.

\bibitem{mesh}  Meshkov A~G, Necessary conditions of the integrability, {\it Inverse Problems}, {\bf 10}, 635--653, 1994.

\bibitem{meshsokHyp}  Meshkov  A~G and  Sokolov V~V,  Hyperbolic equations with symmetries of third order,  {\it  Theoret. and Math. Phys.}, {\bf 166}(1), 43--57, 2011.

\bibitem{meshsok} Meshkov A~G and  Sokolov V~V, Integrable evolution equations with constant separant, {\it Ufa Mathematical Journal}, {\bf 4}(3),  104--154, 2012.

\bibitem{MNW} Mikhailov A~V, Novikov V~S and Wang J~P. On Classification of Integrable Nonevolutionary Equations. {\it Studies in Applied Mathematics}, {\bf 118}(4), 419--457, 2007.

\bibitem{MikSha85} Mikhailov A~V and Shabat A~B,
Integrability conditions for systems of two equations
 $u_t = A(u) u_{xx} {+} B(u,u_x)$ I, 
{\it Theor. Math. Phys.}, {\bf 62}(2), 163--185,  1985.

\bibitem{MikSha86} Mikhailov A~V and Shabat A~B, 
Integrability conditions for systems of two equations
$ u_t = A(u) u_{xx} + B(u, u_x)$ II,
{\it Theor. Math. Phys.}, {\bf 66}(1), 47--65, 1986.

\bibitem{MikShaSok91} Mikhailov A~V, Shabat A~B and Sokolov V~V, 
Symmetry Approach to Classification of Integrable Equations,
{\it What is integrability?} {\rm Ed. V.E. Zakharov},
Springer Series in Nonlinear Dynamics, Springer-Verlag, 115--184, 1991.

\bibitem{MikShaYam87} Mikhailov A~V, Shabat A~B and Yamilov R I, Symmetry approach to classification of nonlinear equations.
Complete lists of integrable systems.
{\it Russian Math. Surveys}, {\bf 42}(4), 1--63, 1987.

\bibitem{MSY} Mikhailov  A~V, Shabat  A~B and  Yamilov R I, Extension of the module of invertible transformations. Classification of integrable systems. {\it Commun. Math. Phys.}, {\bf 115}(1), 1--19, 1988.

\bibitem{miksokcmp}  Mikhailov A~V and  Sokolov V~V, Integrable ODEs
on Associative Algebras, {\it Comm.~in Math. Phys.}, {\bf 211}(1), 231--251, 2000.

\bibitem{int2} Mikhailov A~V and  Sokolov V~V,   Symmetries of differential equations and the problem of Integrability, {\it Integrability}, Ed. Mikhailov A V, {\em Lecture Notes in Physics}, 
Springer, {\bf 767}, 19--88, 2009.

\bibitem{mokfer}  Mokhov O~I and Ferapontov E~V, Nonlocal Hamiltonian operators of hydrodynamic type associated with constant curvature metrics, {\it Russian Math. Surveys}, {\bf 45}(3), 218--219, 1990.

\bibitem{NovWang} Novikov V~S and Wang J~P, Symmetry structure of integrable nonevolutionary equations. {\it Studies in Applied Mathematics}, {\bf 119}(4), 393--428, 2007.

\bibitem{odrubsok} Odesskii A~V, Roubtsov V~N and Sokolov V~V, Bi-Hamiltonian ODEs with matrix variables, {\it Theoret.~and Math.~Phys.}, {\bf 171}(1), 442--447,  2012.

\bibitem{Olv93} Olver P~J, {\it Applications of Lie Groups
to Differential Equations}, (2nd edn), Graduate Texts in Mathematics, {\bf 107}, Springer-Verlag, New York,  1993.

\bibitem{OlvSok98} Olver P~J and Sokolov V~V,
Integrable Evolution Equations on Associative Algebras,
{\it Commun.  Math. Phys.}, {\bf 193}, 245--268, 1998.

\bibitem{ow} Olver P~J and Wang J~P, Classification of integrable one-component systems on associative algebras, {\it Proc. London Math. Soc.}, {\bf 81}(3), 566--586, 2000.

\bibitem{ore} Ore O, Theory of non-commutative polynomials, {\it Ann. Math.}, {\bf 34}, 480--508,  1933.

\bibitem{sw}  Sanders J~A and Wang J~P, On the Integrability of Homogeneous Scalar Evolution Equations,
{\it J. Differential Equations}, {\bf 147}, 410--434, 1998.

\bibitem{SawKot74} Sawada S and Kotera T, 
A method for finding $N$-soliton solutions of the KdV and KdV-like equation,  {\it Prog. Theor. Phys.}, {\bf 51}, 1355--1367, 1974.

\bibitem{sokkn} Sokolov  V~V, Hamiltonian property of the Krichever-Novikov equation, {\it Sov. Math. Dokl.}, {\bf 30}, 44--47,  1984.

\bibitem{sok1} Sokolov  V~V, On the symmetries of evolution equations, {\it Russ. Math. Surv.}, {\bf 43}(5), 165--204, 1988.


\bibitem{sokshab} Sokolov V~V and Shabat A~B, Classification of integrable evolution equations, {\it Soviet Scientific Reviews},  Section C, {\bf 4},  221--280, 1984.

\bibitem{soksvin}  Sokolov V~V and Svinolupov S~I, Weak nonlocalities in evolution equations,  {\it  Math. Notes} {\bf 48}(6), 
1234--1239, 1990.

\bibitem{SokWol99} Sokolov V~V and Wolf T,
A symmetry test for quasilinear coupled systems,
{\it Inverse Problems}, {\bf 15}, L5--L11, 1999.


\bibitem{svin4}  Svinolupov S~I, Second-order evolution  equations with symmetries,
{\it Russian Mathematical Surveys}, {\bf 40}(5), 241--242, 1985.


\bibitem{soksvin1} Svinolupov S~I and Sokolov V~V, Evolution equations with nontrivial conservative laws, {\it Funct. anal. appl.}, {\bf 16}(4),  317--319, 1982.

 \bibitem{soksvin2}  Svinolupov S~I and Sokolov V~V, On conservations laws for the equations possessing nontrivial Lie-B\"acklund algebra, {\it Integrable systems: collection of the papers}. Ed. Shabat A~B, BB AS USSR, Ufa., 53--67, (in Russian), 1982.


\bibitem{SvSok26} Svinolupov S~I and  Sokolov V~V, Factorization of evolution equations, {\it Russian Math. Surveys}, {\bf 47}(3), 
127--162, 1992.

\bibitem{cvsokyam} Svinolupov S~I, Sokolov V~V and Yamilov R I,  On B\"acklund transformations for integrable evolution equations, {\it Sov. Math., Dokl.}, {\bf 28}, 165--168, 1983.

\bibitem{wolf} Wolf~T, The program CRACK for solving PDEs in general relativity, {\it Relativity and Scientific Computing}, 
{\rm Ed. F. W. Hehl, R. A. Puntigam and H. Ruder}, Berlin: Springer, 241--257, 1996.

\bibitem{zib} Zhiber A~V, Quasilinear hyperbolic equations with an infinite-dimensional symmetry algebra,  {\it Russ AC SC Izv. Math.}, {\bf 45}(1), 33--54, 1995.

\bibitem{zibshab1} Zhiber A~V and Shabat A~B, Klein-Gordon equations with a nontrivial group,  {\it Sov. Phys. Dokl.}, {\bf 24}(8), 608--609, 1979.

\bibitem{zibsok} Zhiber A~V and Sokolov V~V, Exactly integrable hyperbolic equations of Liouville type,  {\it Russ. Math. Surv.} {\bf 56}(1), 61--101, 2001.

\bibitem{sokwol3} Sokolov V~V and Wolf T,  On non-abelization of integrable polynomial ODEs, to appear, 
\quad arXiv nlin. 1809.03030, \, 1807.05583.

\bibitem{WoEf} Wolf T and Efimovskaya O,  On integrability of the Kontsevich non-abelian ODE system, Lett. in Math. Phys.,  
{\bf 100}(2), 161--170, 2012,  \quad arXiv:1108.4208v1 [nlin.SI]



\end{thebibliography}
\end{document}